\documentclass[useAMS,a4paper,fleqn,usenatbib]{mnras}
\usepackage{color}
\usepackage{graphicx}	
\usepackage{amsmath}	
\usepackage{amssymb}	
\usepackage{multicol}        
\usepackage{bm}		
\usepackage{pdflscape}	
\usepackage{comment}
\usepackage[T1]{fontenc}
\usepackage{ae,aecompl}
\usepackage[latin1]{inputenc}

\usepackage{txfonts}

\newcommand{\xmm}{{\em XMM--Newton}}
\newcommand{\nus}{{\em NuSTAR}}
\newcommand{\chandra}{{\em Chandra}}
\newcommand{\suz}{{\em Suzaku}}
\newcommand{\integral}{{\em INTEGRAL}}

\newcommand{\swift}{{\em Swift}}

\newcommand{\fermi}{{\em Fermi}}
\newcommand{\vla}{{\em VLA}}
\newcommand{\atca}{{\em ATCA}}
\newcommand{\gmrt}{{\em GMRT}}

\newcommand{\aer}[3]{$#1^{+ #2}_{- #3}$}
\newcommand{\aerm}[3]{#1^{+ #2}_{- #3}}

\newcommand{\ser}[2]{$#1 \pm #2$}
\newcommand{\serm}[2]{#1 \pm #2}
\newcommand{\serexp}[3]{($#1 \pm #2) \times 10^{#3}$}
\newcommand{\lsup}[1]{$< #1$}
\newcommand{\linf}[1]{$> #1$}
\newcommand{\expo}[2]{$ #1 \times 10^{#2}$}
\newcommand{\tento}[1]{$10^{#1}$}
\newcommand{\tentom}[1]{10^{#1}}
\newcommand{\expom}[2]{ #1 \times 10^{#2}}

\newcommand{\chisq}{\chi^{2}}
\newcommand{\rchisq}{\chi^{2}/\textrm{dof}}
\newcommand{\dchi}{\Delta \chi^{2}}
\newcommand{\ddof}{\Delta \textrm{dof}}

\newcommand{\cut}{E_{\textrm{c}}}
\newcommand{\nh}{N_{\textrm{H}}}

\newcommand{\mbh}{M_{\textrm{BH}}}
\newcommand{\msun}{$M_{\odot}$}

\newcommand{\lbol}{L_{\textrm{bol}}}

\newcommand{\fek}{Fe~K$\alpha$}

\newcommand{\hi}{H\,{\sc i}}

\newcommand{\xspec}{{\sc xspec}}

\newcommand{\pexrav}{{\sc pexrav}}

\newcommand{\zphabs}{{\sc zphabs}}
\newcommand{\mekal}{{\sc mekal}}

\newcommand{\gammax}{\Gamma_{\textrm{X}}}

\newcommand{\cflux}{S_{\textrm{1.4GHz}}^{\textrm{core}}}
\newcommand{\lflux}{S_{\textrm{1.4GHz}}^{\textrm{lobes}}}
\newcommand{\clum}{L_{\textrm{1.4GHz}}^{\textrm{core}}}
\newcommand{\llum}{L_{\textrm{1.4GHz}}^{\textrm{lobes}}}

\newcommand{\xlum}{L_{\textrm{2--10keV}}}

\newcommand{\lbolr}{L_{\textrm{bol}}^{\textrm{radio}}}
\newcommand{\lbolx}{L_{\textrm{bol}}^{\textrm{X}}}

\newcommand{\edd}{\epsilon_{\textrm{Edd}}}

\newcommand{\xirx}{\xi_{\textrm{RX}}}
\newcommand{\xirm}{\xi_{\textrm{RM}}}

\newcommand{\cpn}{C_{\textrm{pn}}}
\newcommand{\cbat}{C_{\textrm{BAT}}}
\newcommand{\cibis}{C_{\textrm{IBIS}}}

\newcommand{\fluxcgs}{ergs~s$^{-1}$~cm$^{-2}$}
\newcommand{\lumcgs}{ergs~s$^{-1}$}
\newcommand{\kms}{km~s$^{-1}$}

\newcommand{\sqcm}{cm$^{-2}$}

\newcommand{\refl}{\mathcal{R}}
\newcommand{\ew}{EW_{K\alpha}}

\newcommand{\zero}{0318+684}
\newcommand{\pksz}{PKS~0707-35}
\newcommand{\mrk}{Mrk~1498}
\newcommand{\pksa}{PKS~2331-240}
\newcommand{\pksb}{PKS~2356-61}

\usepackage[space]{grffile}
\usepackage[normalem]{ulem}
\usepackage{hyperref}
\usepackage{comment}
\bibpunct{(}{)}{;}{a}{}{,}

\begin{document}
\title[X-ray properties of giant radio galaxies]{Hard X-ray selected giant radio galaxies -- I. The X-ray properties and radio connection}

\author[F. Ursini]
{F. Ursini,$^{1}$\thanks{e-mail: \href{mailto:francesco.ursini@inaf.it}{\texttt{francesco.ursini@inaf.it}}} 
	L. Bassani,$^{1}$
	F. Panessa,$^{2}$
	A.~J. Bird,$^{3}$
	G. Bruni,$^2$
	M. Fiocchi,$^2$
	A. Malizia,$^{1}$
	\newauthor
	L. Saripalli,$^{4}$
	and
	P. Ubertini$^{2}$
	\\
	$^1$ INAF-Osservatorio di astrofisica e scienza dello spazio di Bologna, Via Piero Gobetti 93/3, 40129 Bologna, Italy. \\
	$^2$ INAF-Istituto di Astrofisica e Planetologia Spaziali, via Fosso del Cavaliere 100, 00133 Roma, Italy.\\
	$^3$ School of Physics and Astronomy, University of Southampton, SO17 1BJ, UK.\\
	$^{4}$ Raman Research Institute, C. V. Raman Avenue, Sadashivanagar, Bangalore 560080, India.
}

\date{Released Xxxx Xxxxx XX}


\maketitle

\label{firstpage}

\begin{abstract}
We present the first broad-band X-ray study of the nuclei of 14 hard X-ray selected giant radio galaxies, based both on the literature and on the analysis of archival X-ray data from \nus, \xmm, \swift\ and \integral. The X-ray properties of the sources are consistent with an accretion-related X-ray emission, likely originating from an X-ray corona coupled to a radiatively efficient accretion flow. We find a correlation between the X-ray luminosity and the radio core luminosity, consistent with that expected for AGNs powered by efficient accretion.
	In most sources, the luminosity of the radio lobes and the estimated jet power are relatively low compared with the nuclear X-ray emission.
This indicates that either the nucleus is more powerful than in the past, consistent with a restarting of the central engine, or that the giant lobes are dimmer due to expansion losses.
\end{abstract}


\section{Introduction}\label{sec:intro}
Active galactic nuclei (AGNs) are powerful sources emitting across all the electromagnetic spectrum. Their central engine is thought to be a supermassive black hole surrounded by an accretion disc, mostly emitting in the optical/UV band. The X-ray emission is thought to originate, at least in radio-quiet AGNs, via Comptonization of disc photons in a hot corona \cite[e.g.][]{haardt&maraschi1991,hmg1994,hmg1997}. 
The primary X-ray emission can be absorbed by the circumnuclear material, and also Compton reflected by the disc \cite[e.g.][]{george&fabian1991,MPP1991} or the torus at pc scales \cite[e.g.][]{ghm1994,matt2003}.
In radio galaxies, powerful jets are also observed, producing radio through gamma radiation via synchrotron and synchrotron self-Compton mechanisms.
From the seminal work of \cite{fanaroff&riley}, radio galaxies are divided into two subclasses according to their morphology and radio power: the
low-power Fanaroff-Riley (FR) I and the high-power FR II. 
FR Is are more compact and their radio luminosity peaks near the nucleus, while FR IIs exhibit well-separated radio lobes with bright outer edges. When jets are observed, they are more collimated in FR IIs. 
Also the origin of the X-ray emission is likely different between the two classes, being mostly related to the jet in FR Is \cite[e.g.][]{balmaverde2006,hardcastle2009,mingo2014} and to the accretion disc/corona system in FR IIs \cite[e.g.][]{grandi2006}.

Giant radio galaxies (GRGs) are the largest (linear size $>0.7$ Mpc) 
single objects in the Universe, and represent an extreme class among radio-loud AGNs. Their extraordinary size, extending well beyond the host galaxy, makes them ideal targets to study the large-scale structure of the Universe and probe the warm-hot intergalactic medium \cite[][]{malarecki2013,malarecki2015}. 
Moreover, according to the evolution models of radio galaxies \cite[e.g.][]{kaiser1999,hardcastle2013}, GRGs should be very old and/or residing in a very low-density environment compared with regular radio galaxies \cite[e.g.][]{subra1996,mack1998,machalski2004,malarecki2015}. The origin and growth of GRGs is still an open issue, and could be related to the restarting of their central engines during multiple activity phases \cite[e.g.][]{subra1996}. Also the environment can be an important factor, since the radio lobes tend to grow to giant sizes towards low-density regions \cite[]{malarecki2015}; in particular, the radio axes of GRGs are preferentially closer to the minor axes of the host galaxies \cite[]{saripalli2009}. 

So far, around 300 GRGs have been reported in the literature, mostly from radio surveys \cite[][and references therein]{wez2016}. Their discovery is challenging because, despite their huge total energy content, their large volume implies a very low energy density and a very low surface brightness, thus requiring sensitive radio telescopes. Another problem is the large angular size, which makes it difficult to distinguish the real radio structure from physically unrelated sources. Indeed, in the complete sample of 3CR extragalactic radio sources \cite[]{laing1983}, only around 6 per cent of them are found to be giant, and this fraction is only 1 per cent at redshift $z<1$ \cite[]{ishwara1999}. 
Among the radio galaxies of the more sensitive Australia Telescope Low Brightness Survey Extended Sources Sample, around 12 per cent are candidate giants, with about 3 per cent being giants at $z<1$ \cite[]{saripalli2012}.

Recently, \cite{bassani2016} extracted a sample of 64 AGNs with extended radio morphology from the hard X-ray catalogues of \integral/IBIS \cite[]{malizia2012} and \swift/BAT \cite[]{bat70}. Surprisingly, the fraction of GRGs was found to be 22 per cent, i.e. significantly larger than what is generally found in radio surveys. Also, 60 per cent of objects were found to have linear sizes above 0.4 Mpc. It is still unclear why the hard X-ray selection favours the detection of giant radio sources. This could be partly due to observational biases that prevent the detection of GRGs in radio surveys. However, hard X-ray selected AGNs are also biased towards luminous and highly accreting sources; this could in turn suggest that hard X-ray selected radio galaxies are those with a central engine powerful enough to produce giant radio structures \cite[for a detailed discussion, see][]{bassani2016}. 

In this paper, we discuss for the first time the broad-band X-ray properties of a sample of hard X-ray selected GRGs, and more specifically of the 14 GRGs selected in \cite{bassani2016}, and investigate their connection with the radio emission. 
To this aim, we gathered information either from the literature or by direct analysis of archival X-ray data, focusing in particular on \nus. A study of the radio morphology and evolution of these sources will be presented in a forthcoming work (Bruni et al., in preparation). The structure of the paper is as follows. In Section \ref{sec:sample} we present the sample of hard X-ray selected GRGs. In Section \ref{sec:obs} we present the X-ray observations analysed in this work. 
We discuss the results in Section \ref{sec:discussion}, and summarize the conclusions in Section \ref{sec:summary}.

\section{The sample}\label{sec:sample}
We collected X-ray data for all the 14 radio galaxies reported as giants in \cite{bassani2016}, namely showing a linear extent above 0.7 Mpc (for $H_0=71$ \kms~Mpc$^{-1}$, $\Omega_M=0.27$, $\Omega_{\Lambda}=0.73$). All these sources display a FR II radio morphology.  
The sample includes local AGNs ($z \lesssim 0.2$) of different optical classes, type 1 and 2 objects being equally represented (see Tables \ref{tab:log2} and \ref{tab:log3}). 
We also collected the available information on the radio fluxes from the literature, mostly relying on high-resolution (5--45 arcsec) 1.4 GHz data that allowed disentangling the different contributions from the core and the lobes. 
Concerning the X-ray data, we focused on broad-band spectra including \integral/IBIS and/or \swift/BAT at high energies (14--195 keV), and \xmm, \chandra\ or \suz\ at lower energies (0.3--10 keV). For 3 sources out of 14, only \swift/XRT data are available, and have been analysed in \cite{panessa2016}. As of April 2018, 7 sources have been observed by \nus\ in targeted observations, whose data either are published or will be discussed in the following. The basic data of the 5 sources analysed in the present work, together with their radio observations, are reported in Table \ref{tab:log2}. The data of the 9 sources with X-ray information taken from the literature are reported in Table \ref{tab:log3}. 
The black hole masses were collected by \cite{panessa2016}, to whom we refer the reader for the details; the mass of \pksa\ was estimated from the stellar velocity dispersion by \cite{lorena2017}. 

\begin{table*}
	\begin{center}
		\caption{ \label{tab:log2} The giant radio galaxies analysed in this work, with the list of X-ray data used here and radio data with corresponding references. 
			}
		\begin{tabular}{ l c c c c c c c } 
			\hline 
Name&$z$&Optical& Galactic $\nh$ & $\log \mbh$&X-ray&Radio&Ref.\\
&&class& (\tento{20} \sqcm)& &data&data&\\ \hline
0318+684&0.090100&Sy1.9&30.8&-&\textit{XMM+NuSTAR}+BAT+IBIS&\vla&\cite{lara2001}\\[2pt]
PKS~0707-35&0.110800&Sy2&17.0&-&\nus+BAT&\atca&\cite{saripalli2013}\\[2pt]
Mrk~1498&0.054700&Sy1.9&1.8&8.59& \textit{XMM+NuSTAR}+BAT&\vla&\cite{schoenmakers2000} \\[2pt]
PKS~2331-240&0.047700&Sy1.8&1.6&8.75&\textit{XMM+NuSTAR}+BAT&\vla & \cite{lorena2017} \\[2pt]
PKS~2356-61&0.096306&Sy2&1.5&8.96&\nus+BAT&\atca&\cite{subra1996}\\
\hline 	
		\end{tabular}
	\end{center}
\end{table*} 

\begin{table*}
	\begin{center}
		\caption{ \label{tab:log3} The giant radio galaxies with X-ray data taken from the literature, with radio data and corresponding references. 
			}
		\begin{tabular}{ l c c c c c c c } 
			\hline 
Name&$z$&Optical&$\log \mbh$&X-ray&Ref.&Radio&Ref.  \\
&&class&&data&&data& \\ \hline
B3~0309+411b&0.134000&Sy1&-&\textit{XMM}+BAT+IBIS&\cite{molina2008}&\vla&\cite{schoenmakers2000}  \\[2pt]
4C~73.08&0.058100&Sy2&-&\textit{XMM+NuSTAR}&\cite{ctrg}&\vla&\cite{lara2001}\\[2pt]
HE~1434-1600&0.144537&BLQSO&8.64&\swift/XRT+BAT&\cite{panessa2016}&\vla&\cite{letawe2004}  \\[2pt]
IGR~J14488-4008&0.123&Sy1.5&8.58&\textit{XMM}+BAT+IBIS&\cite{molina_14488}&\gmrt&\cite{molina_14488} \\[2pt]
4C~63.22&0.20400&Sy1&-&\swift/XRT+BAT&\cite{panessa2016}&\vla&\cite{lara2001} \\[2pt]
4C~34.47&0.20600&Sy1&8.01&\textit{XMM}+BAT&\cite{ricci2017}&\textit{WSRT}&\cite{jagers1982} \\[2pt]
IGR~J17488-2338&0.240&Sy1.2&9.11&\textit{XMM}+IBIS&\cite{molina_17488}&\vla&\cite{molina_17488}  \\[2pt]
PKS~2014-55&0.060629&Sy2&-&\swift/XRT+BAT&\cite{panessa2016}&\atca&\cite{saripalli2007} \\[2pt]
4C~74.26&0.10400&Sy1&9.37&\textit{XMM}+BAT+IBIS&\cite{molina2008}&\vla&\cite{lara2001} \\
&&&&\nus+XRT&\cite{lohfink2017}&&\\
\hline
		\end{tabular}
	\end{center}
\end{table*} 

\section{\nus\ observations, data reduction and analysis}\label{sec:obs}
\nus\ \cite[][]{harrison2013nustar} observed \zero, \pksz, \mrk, \pksa\ and \pksb\ as part of the \nus\ extragalactic survey. 
\subsection{The sources}

\begin{table}
	\begin{center}
		\caption{\label{tab:nustar} Logs of the \textit{NuSTAR} observations analysed in this work.}
		\begin{tabular}{l c c c}
			\hline
			Source & Obs. Id. & Start time (UTC)  & Net exp.\\ 
			& & yyyy-mm-dd & (ks)  \\ 
			\hline
			\zero &60061342002 &2016-05-04& 24 \\[2pt]
			\pksz &60160285002&2016-11-24&19\\[2pt]
			\mrk & 60160640002&2015-05-11&24\\[2pt]
			\pksa &60160832002&2015-07-30 &21\\[2pt]
			\pksb&60061330002&2014-08-10&23\\
			\hline
		\end{tabular}
	\end{center}
\end{table}

\zero\ is among the largest radio sources of our sample, having a linear size of 1.5 Mpc \cite[]{bassani2016}. It hosts an AGN optically classified as a Seyfert 1.9 \cite[]{veron2006}. This source was also observed by \xmm\ in 2006 and studied by \cite{winter2008}, who reported an X-ray absorbing column density of around \expo{4}{22} \sqcm\ and an X-ray photon index $\gammax\simeq 1.5$.

\pksz\ has been classified as a type 2 source by \cite{tazaki2013} from the X-ray obscuration ($\nh > \tentom{22}$ \sqcm) observed with \suz. \cite{tazaki2013} also reported $\gammax\simeq 1.66$ and found the presence of a relatively weak reflection component compared with typical Seyfert galaxies.
The radio morphology is complex, as it shows two distinct lobe pairs, offset from each other at an angle of 30 deg, indicating a restarting activity scenario \cite[]{saripalli2013}.

\mrk\ is a peculiar source, optically classified as a Seyfert 1.9 \cite[]{veron2006}, showing large-scale ionization cones in the optical band that are not aligned with the radio lobes, and that might be interpreted as due to a fading or obscured AGN \cite[]{keel2015,keel2017}. 
From \suz\ data, \cite{eguchi2009} found an X-ray absorbing column density of a few $\times \tentom{23}$ \sqcm\ and $\gammax\simeq 1.8$, and suggested the obscuration is produced by a patchy torus.
A multiwavelength study on this source will be presented in a forthcoming work (Hernandez-Garc\'ia et al., in prep.). Here we focus on the \nus\ data, complemented by a 2007 \xmm\ observation.

\pksa\ has a peculiar optical spectrum with variable broad emission lines.
This source has been optically classified as a Seyfert 2 \cite[]{parisi2012} and later re-classified as a Seyfert 1.9 \cite[]{lorena2017}, changing to type 1.8 in one year \cite[]{lorena2018}. The radio morphology shows two giant lobes plus a blazar-like core in the center, suggesting the jet has changed its orientation and is now pointing in our line of sight \cite[]{lorena2017}. 
Concerning the X-ray spectrum, \cite{panessa2016} reported $\gammax \simeq 1.70$ and no intrinsic absorption from \swift/XRT+BAT data \cite[see also][]{lorena2018}. This object was also observed once by \nus\ and twice by \xmm\ in 2015, although never simultaneously.
\cite{lorena2017} reported $\gammax \simeq 1.77$ and no intrinsic absorption from the \xmm\ data.

Finally, \pksb\ is a Seyfert 2 \cite[]{veron2006} and a very powerful radio galaxy, with large and bright hotspots \cite[e.g.][]{subra1996,mingo2014,mingo2017}. From \chandra\ data, \cite{mingo2014} concluded that the X-ray spectrum is dominated by an accretion-related continuum, absorbed by a column density of around \expo{1.5}{23} \sqcm. From the same data, \cite{mingo2017} detected synchrotron X-ray emission from one of the hotspots. 
\subsection{The X-ray data}
We report in Table \ref{tab:nustar} the logs of the archival \nus\ data sets of the sources analysed here. 
Since \nus\ is sensitive in the 3--79 keV band, it allows us to constrain the different X-ray spectral components, namely the primary power law-like continuum, the Compton-reflection component producing a bump at 20-30 keV and the \fek\ line at 6.4 keV. 
We included for all sources the \swift/BAT spectra \cite[]{bat} to extend the coverage up to 195 keV, and also \integral/IBIS data \cite[]{ibis} for \zero. We also included the soft X-ray spectra by \xmm\ \cite[][]{xmm} for \zero, \mrk\ and \pksa, to better constrain the absorbing column density and the \fek\ line.

The \nus\ data were reduced using the standard pipeline (\textsc{nupipeline}) in the \nus\ Data Analysis Software (\textsc{nustardas}, v1.8.0), using calibration files from \nus\ {\sc caldb} v20180312. We extracted the spectra 
using the standard tool {\sc nuproducts} for each of the two hard X-ray detectors, which reside in the corresponding focal plane modules A and B (FPMA and FPMB). We extracted the source data from circular regions with a radius of 75 arcsec, and the background from a blank area close to the source. The spectra were binned to have a signal-to-noise ratio larger than 3 in each spectral channel, and not to oversample the instrumental resolution by a factor greater than 2.5. The spectra from the two detectors were analysed jointly, but not combined.

The \xmm\ data were processed using the \xmm\ Science Analysis System (\textsc{sas} v16.1). 
For simplicity, we used EPIC-pn data only, because of the much larger effective area compared with the MOS detectors and to avoid uncertainties due to cross-calibration issues.
The source extraction radii and screening for high-background intervals were determined through an iterative process that maximizes the signal-to-noise ratio \cite[][]{pico2004}. We extracted the background from circular regions with a radius of 50 arcsec, while the source extraction radii were in the range 20--40 arcsec. We binned the spectra to have at least 30 counts per spectral bin, and not oversampling the instrumental resolution by a factor greater than 3.

The \integral\ spectrum of \zero\ consists of ISGRI data from several pointings between revolution 12 and 530 \cite[from the fourth IBIS catalogue;][]{bird2010}. The data extraction was carried out following the procedure described in \cite{molina2013}.

Finally, we included the average 105-month \swift/BAT spectra from the most recent hard X-ray survey \cite[][]{bat105}\footnote{\url{https://swift.gsfc.nasa.gov/results/bs105mon/}}.

\subsection{Spectral analysis}\label{sec:analysis}
The spectral analysis was carried out with the \xspec\ 12.10.0 package \cite[][]{arnaud1996}, using the $\chisq$ minimisation technique. 
All errors are quoted at the 90 per cent confidence level.
We fitted the 3--79 keV \nus\ spectra and the 14--195 keV \swift/BAT spectra simultaneously, leaving the cross-calibration constant $\cbat$ free to vary, after checking for consistency between the two instruments. We did the same for the 20--100 keV \integral/IBIS spectrum and included a cross-calibration constant $\cibis$.
We also included a cross-calibration factor between \nus/FPMA and FPMB, which is always consistent with unity and smaller than 1.02. 
The \xmm/pn data allowed us to extend the analysis down to 0.3 keV for \zero, \mrk\ and \pksa. In these cases, we included a cross-calibration constant $\cpn$ free to vary to account for flux variability between the different observations.

In our fits, we always included Galactic absorption, fixing the hydrogen column densities to the values obtained from the \hi\ map of \cite{kalberla2005}, as given by the tool \textsc{nh} in the \textsc{HEASoft} package.
	The values of Galactic $\nh$ are reported in Table \ref{tab:log2}.
For all models, we adopted the chemical abundances of \cite{angr} and the photoelectric absorption cross-sections of \cite{vern}. 

We first fitted the data with a model consisting of an absorbed power law plus a reflection component and a narrow Gaussian emission line. We used the \pexrav\ model \cite[]{pexrav} in \xspec, which includes the continuum reflected off a neutral medium of infinite column density in a slab geometry. We always fixed the inclination angle at 60 deg, and we assumed solar abundances. We left free to vary the photon index $\gammax$, the reflection fraction $\refl$ and the high-energy cut-off $\cut$ in \pexrav. 
Since the energy of the Gaussian line was always poorly constrained, we fixed it at 6.4 keV (rest-frame), i.e. that expected for neutral \fek\ emission.
In \xspec\ notation, this baseline model reads: \textsc{const*phabs*zphabs*(pexrav + zgauss)}, where \textsc{const} is the cross-calibration constant, \textsc{phabs} is the fixed Galactic absorption, \textsc{zphabs} is the (redshifted) intrinsic absorption. This model is designated as Model A in the following.

When including pn data down to 0.3 keV, we noted a ``soft excess'' on top of the absorbed power law, which is commonly observed in the X-ray spectra of type 2 AGNs \cite[e.g.][]{turner1997,gmp2005}. 
	In radio-quiet Seyfert 2s, this excess is generally explained as optically-thin scattered continuum \cite[e.g.][]{turner1997,ueda2007} and/or photoionized emission from circumnuclear gas \cite[e.g.][]{gb2007}. In obscured radio galaxies, on the other hand, the excess can be attributed to unabsorbed X-ray emission from the jet \cite[e.g.][]{hardcastle2009,mingo2014}. In any case, the excess is generally well described by a power law.
	We thus included to the baseline model (Model A) a secondary, unabsorbed power law. In \xspec\ notation, this model (Model B) reads: 
	\textsc{const*phabs*[zphabs*(pexrav + zgauss)+const2*powerlaw]}.
	If the soft excess is interpreted as a scattered component, then \textsc{const2} would represent the scattered fraction $f_s$, which is generally of a few per cent or less \cite[e.g.][]{turner1997}. The parameters of the second power law are tied to those of \textsc{pexrav}. 
	Finally,	
to test the presence of photoionized emission, we added a thermal component to Model B. We used the \mekal\ model in \xspec. Then, in \xspec\ notation, this model (Model C) reads: 
\textsc{const*phabs*[zphabs*(pexrav + zgauss + mekal)+const2*powerlaw]}.

The results obtained for each source are discussed below and summarized in Table \ref{tab:params}, where we report the main best-fitting parameters. The X-ray spectral parameters for the other sources in our sample have been collected from the literature and are reported in Table \ref{tab:literature}.
\subsubsection{0318+684}
\nus\ observed this source in 2016 with a net exposure of 24 ks, while \xmm\ observed the source in 2006 (Obs. Id. 0312190501) with a net exposure of 6.5 ks. As a consistency check, we fitted the \nus\ and \xmm/pn spectra in the common bandpass 3--10 keV with a simple power law. The photon index was found to be \ser{1.26}{0.09} in \nus\ and \ser{1.2}{0.1} in pn, i.e. the spectral shape is consistent with being the same. On the other hand, the normalization measured by pn is a factor of 2 higher than in \nus, indicating a flux variation between the two observations. However, the spectral shape being consistent, we fitted simultaneously the \nus\ and pn spectra, also including \swift/BAT and \integral/IBIS to get the broadest energy band (0.3--195 keV). 

Model A provides a decent fit to the data ($\rchisq=265/238$), but leaves significant residuals in the soft band. The fit is improved using Model B ($\rchisq=245/237$, $\dchi/\ddof = -20/-1$), and we obtain decent constraints on the intrinsic photon index and on the column density, as reported in Table \ref{tab:params}. These results are consistent, within the errors, with those reported by \cite{winter2008} from the analysis of the pn data alone. We only have an upper limit to the reflection fraction $\refl<0.17$, while the Gaussian line at 6.4 keV has an equivalent width of $\ew = \serm{50}{40}$ eV. The high-energy cut-off is found to be $\cut=\aerm{60}{60}{30}$ keV. 

\subsubsection{PKS 0707-35}
Only \nus\ and BAT data were available for this source, we thus fitted the spectra with the baseline Model A. We find a good fit with $\rchisq=100/95$. The cross-calibration constant is large ($\cbat=\serm{3.0}{0.8}$), but leaving the photon index free to change between \nus\ and BAT does not improve the fit. We only derive rough upper limits to the presence of a reflection component ($\refl<2$) and of the \fek\ line ($\ew < 350$ eV). We note that the 2--10 keV flux measured by \nus\ is a factor of 4 less than that by \suz\ (\expo{5}{-12} \fluxcgs) as given in \cite{tazaki2013}.
\subsubsection{Mrk 1498}
This source was observed by \nus\ in 2015 with a net exposure of 24 ks, and by \xmm\ in 2007 (Obs. Id. 0500850501) with a net exposure of 7.5 ks. Since the spectrum shows strong signatures of X-ray absorption, to check for consistency between \nus\ and pn we proceeded as follows. We first fitted the 3--10 keV \nus\ spectra with a simple power law modified by \zphabs. We found a photon index \ser{1.58}{0.13} and a column density of \serexp{19}{3}{22} \sqcm. Then, we fitted the pn spectrum in the same energy band and with the same model, fixing the parameters and only allowing for a free cross-calibration constant. We found a good fit and $\cpn=\serm{0.93}{0.03}$, indicating a good agreement between the two spectra. 
We fitted jointly the \nus, pn and BAT spectra in the 0.3--195 keV band. 
Model A gives a very poor fit ($\rchisq= 763/394$). The fit is greatly improved using Model B ($\rchisq= 428/392$, $\dchi/\ddof=-335/-2$), but with significant residuals in the soft band.
Finally, Model C yields a good fit with no prominent residuals ($\rchisq=396/390$, $\dchi/\ddof=-32/-2$).  We obtain a photon index of \ser{1.5}{0.1}, a column density of \serexp{23}{3}{22} \sqcm, an upper limit to the reflection fraction of 0.35, and a cut-off energy of \aer{80}{50}{20} keV. The \fek\ line has an equivalent width of \ser{70}{30} eV. The temperature of the \mekal\ component is found to be \aer{0.15}{0.04}{0.06} keV, while its normalization is \serexp{9}{4}{-5}. Finally, the scattered fraction is \serexp{1.8}{4}{-2}. 
\subsubsection{PKS 2331-240}
\nus\ observed this source in Jul. 2015 with a net exposure of 24 ks; \xmm\ observed the source twice, in May and Nov. 2015 (Obs. Id. 0760990101 and 0760990201) with a net exposure of 18 and 20 ks, respectively.
We first fitted the \nus\ and pn spectra in the 3--10 keV band with a simple power law, finding a discrepancy between the different spectra. Indeed, we found a photon index of \ser{1.81}{0.06} in \nus, \ser{1.72}{0.05} in the first pn spectrum and \ser{1.64}{0.05} in the second pn spectrum. 
Moreover, the 3--10 keV flux in pn was found to be 20 and 30 per cent less than in \nus. 
However, to constrain $\nh$ and the \fek\ line, we chose to include the pn spectra in our fits, leaving the photon index free to vary among the different observations. Since the BAT photon index was poorly constrained, we tied it to that of \nus.
We obtain a decent fit using Model A ($\rchisq=611/540$), with residuals that can be attributed to noise.
Albeit this source is optically a Sy1.9, it is almost unabsorbed \cite[see also][]{lorena2017} and we only find a column density of \serexp{1.1}{0.3}{20} \sqcm\ in excess of the Galactic one. We find an upper limit to the reflection fraction of 0.3 and a \fek\ line with an equivalent width of \ser{50}{20} eV. The \nus\ photon index for the broad-band (0.3--195 keV) fit is \ser{1.91}{0.07}, while for pn we have \ser{1.75}{0.02} and \ser{1.77}{0.02} \cite[consistent with][]{lorena2017} and a cross-calibration constant of \ser{0.65}{0.07} and \ser{0.54}{0.06}, respectively.

\subsubsection{PKS 2356-61}
We fitted the \nus\ and BAT spectra with Model A, which yields a decent fit ($\rchisq=131/159$). We obtain a photon index of \ser{1.7}{0.3} and a column density of \serexp{1.4}{0.5}{23} \sqcm, in agreement with the results of \cite{mingo2014} from \chandra\ data alone. We only derive upper limits to the presence of a reflection component ($\refl<1.3$) and of a \fek\ line ($\ew<170$ eV).

	\begin{table*}
		\begin{center}
			\caption{ \label{tab:params} Best-fitting parameters of the X-ray spectra analysed in this work. }
			\begin{tabular}{ l c c c c c c c c c c  } 
				\hline 
				Name & Model& $\gammax$ & $N_H$ & $\cut$ & $\refl$ &$\ew$ & $\cpn$ & $\cbat$ & $\cibis$ & $\rchisq$  \\
				&&& (\tento{22} \sqcm) &(keV)&&(eV)&&&&\\
				\hline
				0318+684&B &\ser{1.40}{0.12}&\ser{4.0}{0.3}&\aer{60}{60}{30}&\lsup{0.17}&\ser{50}{40}&\ser{2.0}{0.1}&\ser{0.85}{0.15}&\ser{1.3}{0.6} &245/237\\[2pt]
				
				PKS~0707-35&A & \ser{1.6}{0.2} &\aer{8}{7}{6}&\linf{90}&\lsup{2}&\lsup{350}&-&\ser{3.0}{0.8}&-&100/95 \\[2pt]
				
				Mrk 1498&C &\ser{1.5}{0.1}&\ser{23}{2}&\aer{80}{50}{20}&\lsup{0.35}&\ser{90}{50}&\ser{0.94}{0.04}&\ser{1.00}{0.08}&-&396/390 \\[2pt]
				
				PKS~2331-240&A &\ser{1.91}{0.07}&\ser{0.011}{0.003}&\linf{250}&\lsup{0.3}&\lsup{50}&\ser{0.65}{0.07}&\ser{0.45}{0.12}&-&611/540\\
				&&&&&&&\ser{0.54}{0.06}&&&\\[2pt]
				
				PKS~2356-61&A &\ser{1.7}{0.3}&\ser{14}{5}&\linf{55}&\lsup{1.3}&\lsup{170}&-&\ser{0.9}{0.2}&-&131/159\\
				\hline
			\end{tabular}
		\end{center}
	\end{table*}

	\begin{table*}
		\begin{center}
			\caption{ \label{tab:literature} X-ray spectral parameters of the sources reported in the literature. }
			\begin{tabular}{ l c c c c c c  } 
				\hline 
				Name & $\gammax$ &  $N_H$&$\cut$ & $\refl$ &$\ew$ &  Ref. \\
				&& (\tento{22} \sqcm) &(keV)&&(eV)&\\
				\hline
				B3 0309+411b  &\ser{1.90}{0.08}&-&-&\linf{1.2}&\ser{70}{40}&1\\[2pt] 

				4C~73.08 &\ser{1.61}{0.17}&\ser{40}{8}&-&\lsup{2.2}&\ser{120}{100}&2\\[2pt] 
				
				HE~1434-1600 &\ser{1.72}{0.06}&-&-&-&-&3 \\[2pt]
				
				IGR~J14488-4008 &\aer{1.71}{0.16}{0.17}&\ser{0.17}{0.04}&\aer{67}{227}{36}&1(f)&\aer{93}{35}{34}&4\\[2pt]
				
				4C~63.22 &\ser{1.96}{0.08}&-&-&-&-&3 \\[2pt]
				
				4C~34.47 &\ser{1.98}{0.12}&-&\linf{97}&\aer{2.0}{2.1}{1.6}&-&5 \\[2pt]
				
				IGR~J17488-2338 &\ser{1.37}{0.11}&\aer{1.14}{0.26}{0.23}&-&\lsup{1.8}&\aer{128}{61}{62}&6 \\[2pt]
				
				PKS~2014-55 &\ser{1.86}{0.21}&\aer{32}{12}{9}&-&-&-&3 \\[2pt]
				
				4C~74.26 &1.8--1.9&$\sim0.35^a$&\aer{183}{51}{35}&\ser{1.2}{0.7}&90--200&7,8,9 \\
				\hline

			\end{tabular}
		\end{center}
		\begin{flushleft}
			\textit{Note.} $^a$ Ionized absorber. References: 1. \cite{molina2008}, 2. \cite{ctrg}, 3. \cite{panessa2016}, 4. \cite{molina_14488}, 5. \cite{ricci2017}, 6. \cite{molina_17488}, 7. \cite{molina2008}, \cite{digesu2016}, \cite{lohfink2017}.
		\end{flushleft}
	\end{table*} 

\section{Discussion}\label{sec:discussion}
\subsection{X-ray properties}
The X-ray spectral properties of the hard X-ray selected GRGs of our sample are overall consistent with that of normal-size FR II radio galaxies, and more precisely of `high-excitation' radio galaxies (HERGs) powered by efficient accretion \cite[e.g.][]{hardcastle2006,hardcastle2009,mingo2014}.
The X-ray spectra are generally well described by a power law with a photon index $\gammax$ in the range 1.6--1.9, i.e. typical of average AGNs \cite[e.g.][]{malizia2014}. 
In general, past observations have shown that spectral features such as the Compton hump and the \fek\ line tend to be weaker in radio galaxies, compared with radio-quiet Seyferts
\cite[e.g.][]{wozniak1998,eracle2000,grandi2001,molina2008,walton2013}.
From high-sensitivity measurements obtained in recent years with \nus, a weak or absent reflection bump has been found in 3C~382 \cite[]{ballantyne2014,3c382}, 3C~273 \cite[]{madsen2015}, 3C~390.3 \cite[]{lohfink2015} and Cen~A \cite[]{fuerst2016};
on the other hand, signatures of reflection (neutral or ionized) have been observed in 3C~120 \cite[]{lohfink2013}, Cyg~A \cite[]{reynolds2015} and 4C~74.26 \cite[]{lohfink2017}, which is part of our sample. 
Although our results suggest a trend of weak reflection features, future studies on larger samples will be needed to properly estimate the contribution of the reflection component in radio galaxies. 
Interestingly, \cite{king2017} reported an inverse correlation between the radio Eddington luminosity and the X-ray reflection fraction in a sample of AGNs, both radio-quiet and radio-loud. \cite{king2017} interpreted this result in terms of an outflowing, mildly relativistic corona, whose emission is beamed away from the accretion disc and the surrounding material \cite[e.g.][]{belo1999,malzac2001}. This would be consistent with the X-ray corona being the base of the radio jet, as suggested for X-ray binaries \cite[]{markoff2005} and radio galaxies like 3C~120 \cite[][]{lohfink2013}. 

An exponential high-energy cut-off is measured in 4 sources out of 14, two of which have been analysed here (\zero\ and \mrk). In three cases, we only derived lower limits to the cut-off based on \nus\ data. The presence of a high-energy cut-off is consistent with the X-ray emission originating via thermal Comptonization rather than synchrotron and/or inverse Compton in a jet. We also note that none of our sources is a strong gamma-ray emitter, as they are not detected by \fermi\ \cite[]{acero2015}; to our knowledge, the only exception is B3~0309+411b, for which \cite{hooper2016} reported a 0.1--100 GeV flux $\simeq \expom{2.5}{-12}$ \fluxcgs\ from 85-month \fermi\ data. The lack of a strong gamma-ray emission indicates that a high-energy cut-off is likely present below the MeV band.

All in all, the X-ray emission is consistent with being accretion-related and possibly due to a hot Comptonizing corona, that could be outflowing. 
The jet component could give a significant contribution, at least in the soft X-ray band \cite[][]{hardcastle2006,hardcastle2009}, but is unlikely to dominate the overall emission in most sources \cite[see also][]{mingo2017}. In the case of \pksa, the X-ray emission is consistent with originating from external inverse Compton scattering in the jet \cite[]{lorena2017}, but this is the only source of our sample classified as a blazar.

Finally, the X-ray absorption properties are in agreement with the zeroth-order predictions of unified models of AGNs \cite[e.g.][]{antonucci1993,urry&padovani}, as type 1 and 2 objects tend to be unobscured and obscured, respectively \cite[for a detailed discussion on the full sample of hard X-ray selected radio galaxies, see][]{panessa2016}. 

In Table \ref{tab:radio-x} we report the X-ray (2--10 keV) flux and luminosity of all the 14 GRGs of our sample, and the radio (1.4-GHz) flux density and $\nu L_{\nu}$ luminosity of their core and lobes. The radio flux densities were mostly taken from the literature (see Tables \ref{tab:log2} and \ref{tab:log3}). For IGR~J14488-4008, we extrapolated to 1.4 GHz the \gmrt\ measurements at 325 and 610 MHz by \cite{molina_14488}. For PKS~2014-55 the fluxes of the core and of the extended structures were estimated from the \atca\ map \cite[]{saripalli2007}. For PKS~2356-61, we extrapolated to 1.4 GHz the 2.3-GHz and 5-GHz measurements by \cite{morganti1993,morganti1997}. In the case of \pksz, we reported the fluxes of the inner lobes (a) and outer lobes (b) separately \cite[]{saripalli2013}. A more detailed analysis of the radio properties of the sample will be developed in Bruni et al. (in prep.).
\subsection{Radio--X-ray relation}\label{sec:radiox}
\begin{figure}
	\includegraphics[width=\columnwidth]{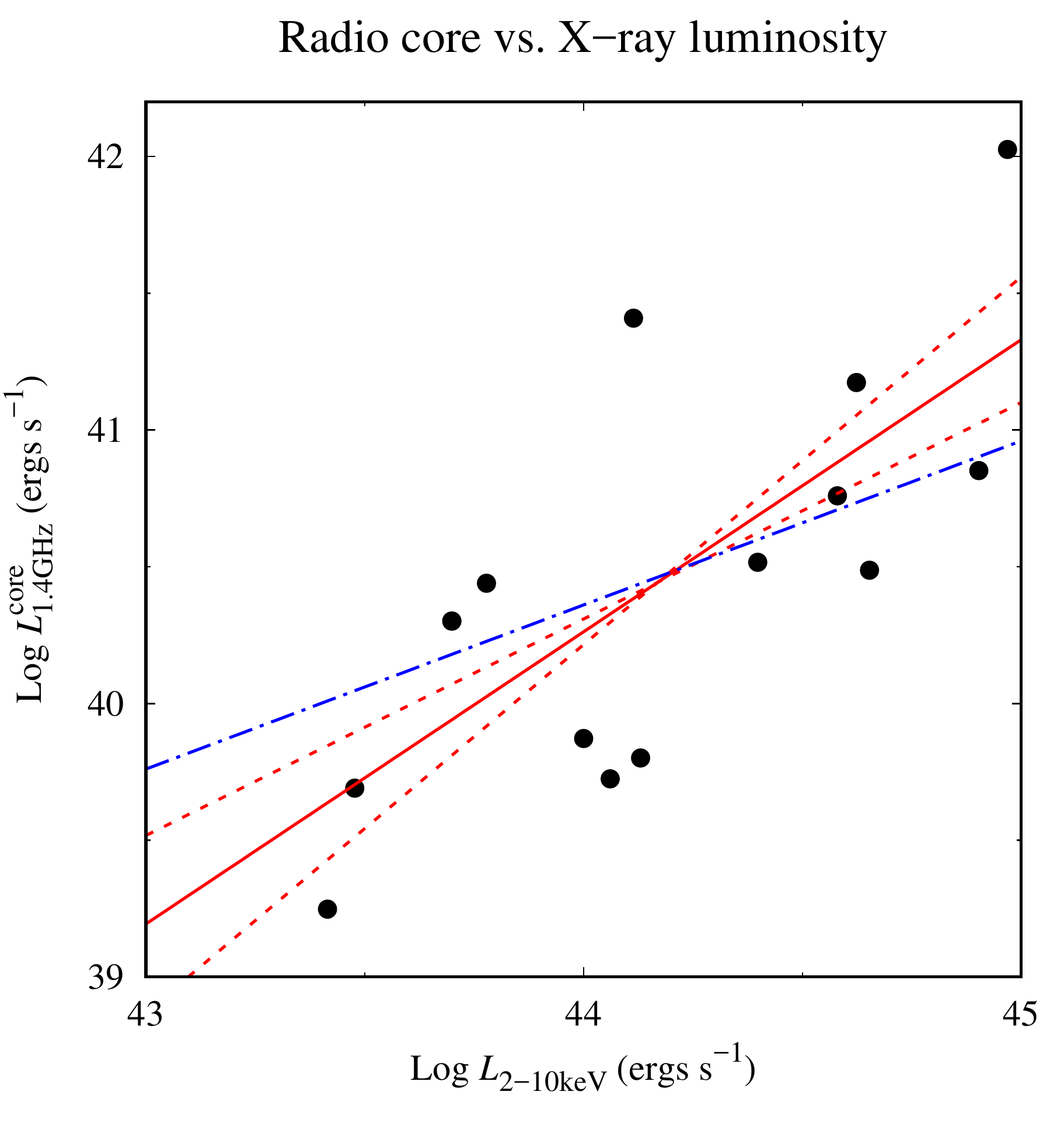}
	\caption{\label{fig:core-x} Radio core 1.4-GHz luminosity versus X-ray 2--10 keV luminosity. The red solid line represents a linear fit in the log-log space, while the red dashed lines correspond to the 90 per cent error on the slope and normalization. The blue dot-dashed line represents the relation of \citet{merloni2003} for a black hole mass of \tento{8} solar masses. }
\end{figure}

A relationship is known to exist between the radio core luminosity, the X-ray luminosity and the black
hole mass in AGNs and X-ray binaries \cite[the so-called fundamental plane of black hole activity:][]{merloni2003,falcke2004,gultekin2009}. 
The original relation of \cite{merloni2003} is:
\begin{equation}
\log L_{\textrm{5GHz}}^{\textrm{core}} = \xirx \log \xlum + \xirm \log M + b_{\textrm{R}}
\end{equation} 
with $\xirx=0.6$, $\xirm=0.78$ and $b_{\textrm{R}}=7.33$.
In fig. \ref{fig:core-x} we plot the radio luminosity $\clum$ against the X-ray luminosity $\xlum$ for our sources. We performed a linear regression fit in the logarithmic space:
\begin{equation}
\log \clum = \xirx \log \xlum +C
\end{equation}
finding $\xirx=\serm{1.1}{0.3}$ and $C=\serm{-7}{12}$. The linear correlation is significant, as we calculate 
a Kendall's coefficient $\tau=0.63$ with a $p$-value of \expo{1}{-3}. However, to properly evaluate the significance of this correlation, we need to take into account the uncertainty on the X-ray and radio luminosities. In principle, both measurement errors and flux variability can contribute to this uncertainty. However, the measurement errors are generally small, i.e. no more than a few per cent; flux variability, on the other hand, can easily amount up to a factor of a few. In particular, the X-ray variability in our sources can be up to 0.3 dex as indicated by the cross-calibration constants of our fits (see Table \ref{tab:params}). Taking 0.3 dex as a fiducial uncertainty on $\xlum$ and $\clum$, we performed a bootstrap on the data, finding a 99 per cent confidence interval for the Kendall's coefficient of $0.15 < \tau < 0.95$. Finally, we tested for a possible distance effect on the correlation, namely the bias introduced by the common dependence of $\xlum$ and $\clum$ on the distance \cite[e.g.][]{merloni2006}. We thus performed a partial Kendall's correlation test among $\clum$ and $\xlum$, using the luminosity distance as the third variable. 
We obtained $\tau=0.42$ with a $p$-value of \expo{4}{-2}, i.e. still a significant correlation. 
Also in this case, we performed a bootstrap on the data, assuming an uncertainty on the distance of 0.4 dex due to the uncertainty on the Hubble flow \cite[]{kording2006}. We obtained a 99 per cent confidence interval for the Kendall's coefficient of $0.15 < \tau < 0.68$.

In fig. \ref{fig:core-x} we overplot the relation of \cite{merloni2003} converting the 5-GHz luminosity into the 1.4-GHz luminosity assuming a radio spectral index $\alpha=0.7$, where $L_{\nu} \propto \nu^{-\alpha}$ \cite[][]{condon2002}, and a black hole mass of \tento{8} solar masses. 
We note that the estimates of the black hole mass are available only in 8 sources out of 14 \cite[i.e. the brightest, see][]{panessa2016}. This uncertainty, coupled with the intrinsic scatter about the fundamental plane, could explain the discrepancy among the data and the relation of \cite{merloni2003}, also given the limited size of our sample. We also note that different estimates of the fundamental plane have been reported, using larger samples \cite[e.g.][]{kording2006,li2008,bonchi2013}. 
However, our results are consistent with previous findings by \cite{panessa2015}, who analysed the 1.4-GHz radio properties of a complete sample of hard X-ray selected AGNs, reporting a steep radio--X-ray correlation ($\xirx \simeq 1$). It is thus possible that the hard X-ray GRGs belong to a different branch in the radio--X-ray correlation. 
The physical meaning of the fundamental plane is the existence of a relationship between accretion power and jet emission \cite[e.g.][]{hardcastle2009}, both in AGNs and in Galactic X-ray binaries \cite[e.g.][]{gallo2003}. Although the underlying mechanism of this coupling is still a matter of speculation, it might suggest that the X-ray emitting corona is the base of the jet \cite[e.g.][]{markoff2005}. 
However, alternative models exist, like the so-called jet emitting discs in X-ray binaries \cite[][]{ferreira2006,pop2010}. Furthermore, X-ray binaries seem to exhibit two different branches in the radio--X-ray correlation: the `standard' branch with $\xirx\simeq 0.6$, and a second branch with $\xirx\simeq 1{-}1.4$ \cite[][]{coriat2011,gallo2012}. The first branch is consistent with the source being powered by a radiatively inefficient accretion flow, in which most of the released energy is advected \cite[][]{adaf} and/or channelled into outflows \cite[][]{adios} or jets \cite[][]{markoff2005}. The second branch is instead consistent with radiatively efficient flows, like the standard \cite{ss1973} accretion disc or the jet-emitting disc \cite[]{ferreira2006}. Extrapolating these results to AGNs, the hard X-ray selected GRGs would be located in the `efficient' branch of the radio--X-ray diagram, since $\xirx\simeq1.1$. 
This in turn suggests that their nuclear activity is driven by a radiatively efficient mode of accretion, with the caveat that several uncertainties remain on the nature of the `efficient' branch, complicating the theoretical interpretation (\citealt{gallo2012}, but see also \citealt{motta2018}). 


\subsection{Bolometric luminosity}
In radio galaxies, the radio luminosity of the lobes is found to be related with the disc luminosity \cite[e.g.][]{willott1999,kording2008,VV2013,vanvelzen2015}. In particular, \cite{vanvelzen2015} reported a linear correlation, in the logarithmic space, between the 1.4-GHz lobes luminosity and the bolometric luminosity estimated from the optical one; the normalization is
\begin{equation}
\log (\llum / \lbol) = -3.57.
\end{equation} 
We used this relation to estimate the bolometric luminosity from $\llum$. This estimate is labelled $\lbolr$ in the following.
The major source of uncertainty on $\lbolr$ is the scatter of the \cite{vanvelzen2015} relation, amounting to 0.47 dex (rms).

The bolometric luminosity can be independently estimated from the X-ray 2--10 keV luminosity using bolometric corrections \cite[e.g.][]{marconi2004,hopkins2007,runnoe2012}. Assuming the \cite{marconi2004} correction, we computed the bolometric luminosity $\lbolx$.
	This quantity is directly related to the present nuclear activity, while $\lbolr$ can be viewed as a tracer of the past activity.
Because the scatter of the \cite{marconi2004} relation is $\sim 0.1$ dex, the main source of uncertainty on $\lbolx$ is the X-ray flux variability, which increases the spread in any bolometric correction \cite[][]{VF2007}. 
We also note that assuming a different bolometric correction would yield estimates of $\lbolx$ within a factor of a few. For example, the linear correction of \cite{runnoe2012} for radio-loud AGNs yields estimates within a factor of $\sim 3$, while the luminosity-dependent correction of \cite{hopkins2007} yields systematically larger estimates of $\lbolx$ up to a factor of 2. We also checked that the estimates of $\lbolx$ are in rough agreement, again within a factor of a few, with the constant correction of \cite{mushotzky2008} for the 14--195 keV luminosity, and with the Eddington ratio-dependent correction of \cite{VF2009}. However, we stress that our main purpose is only to compare the AGN luminosity as traced by the X-ray emission with the luminosity as inferred by the radio lobes. 

\begin{figure}
	\includegraphics[width=\columnwidth]{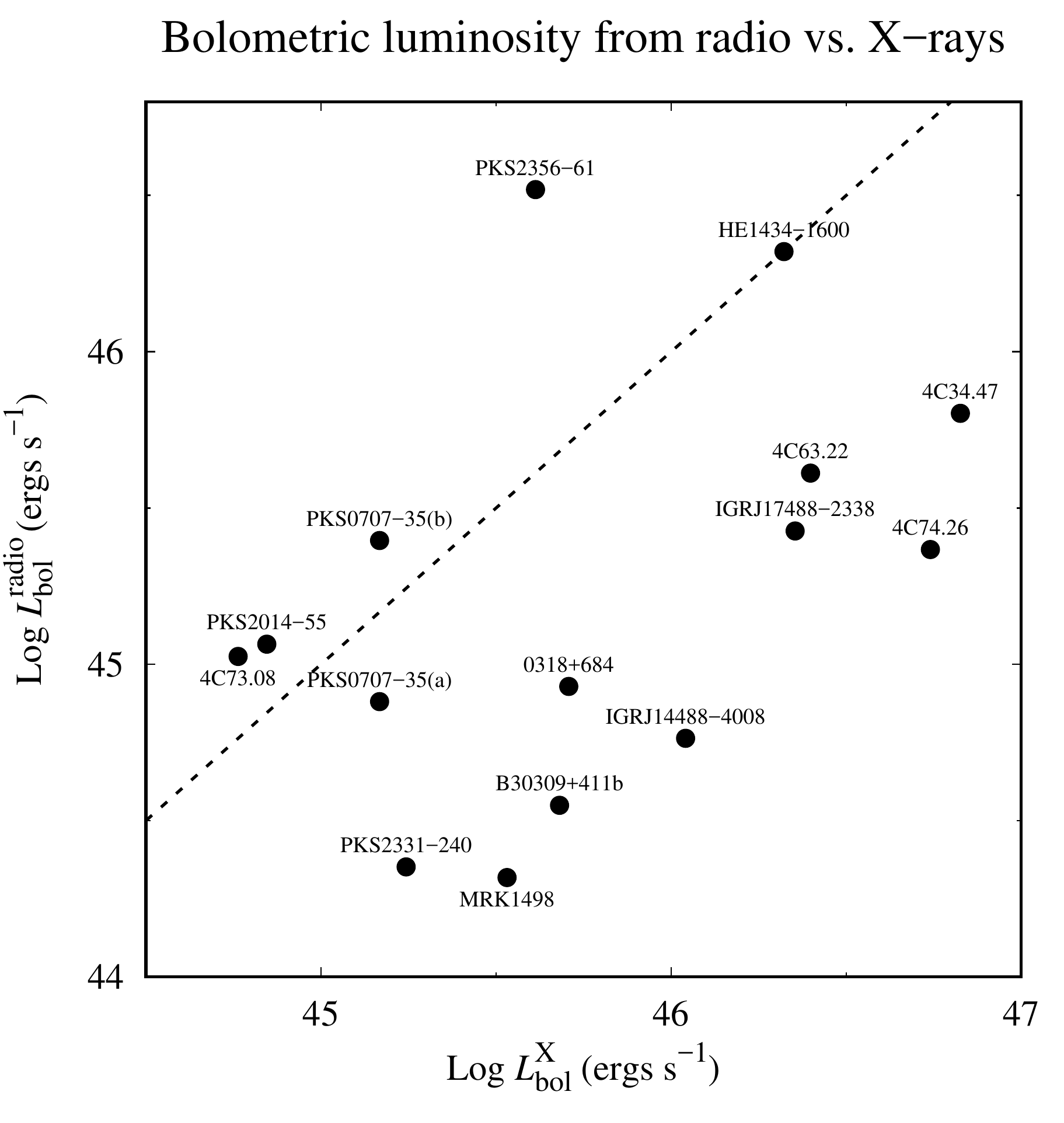}
	\caption{\label{fig:lobes-x} Bolometric luminosity estimated from the radio luminosity of the lobes ($y$-axis) versus that estimated from the 2--10 keV luminosity ($x$-axis). The dashed line represents the identity $y=x$. }
\end{figure}

In Fig. \ref{fig:lobes-x} we plot the two independent estimates of the bolometric luminosity. Most of the objects exhibit a discrepancy between $\lbolr$ and $\lbolx$, $\lbolr$ being around one order of magnitude smaller than $\lbolx$.
Despite all the aforementioned uncertainties, the discrepancy between $\lbolr$ and $\lbolx$ is significant, at least in 9 sources out of 14. 
Interestingly, this discrepancy is consistent with a similar trend seen in  blazars with double radio lobes
\cite[]{pjanka2017}. Indeed, \cite{pjanka2017} found that the blazar jet powers measured from radio lobes are a factor of $\sim 10$ lower than those from blazar-model spectral fitting \cite[]{ghisellini2014} or from the angular shift of the radio core \cite[]{zdziarski2015}. \cite{pjanka2017} proposed two different explanations for this discrepancy. First, it could be a signature of intermittent accretion. In this case, luminous and efficient quasars ($\edd \gtrsim 0.03$) could be caught in a high state lasting only a small fraction of the entire lifetime, which is mostly spent in quiescent states \cite[see also][]{VV2013}. Alternatively, the jet power from the blazar model could be overestimated due to the presence of e$^{\pm}$ pairs \cite[see also][]{sikora2016}. 
In our case, we observe a discrepancy between the nuclear luminosity, traced by the X-rays, and the radio lobes luminosity, estimated from the disc-lobes relationship reported by \cite{vanvelzen2015} for radio selected FR IIs.
	This could mean that the \cite{vanvelzen2015} relationship may not hold for the GRGs in our sample.
	This in turn suggests that either their nuclear luminosity is higher than the average luminosity during their lifetime, or that their lobes luminosity is lower than expected (possibly due to radiative losses).
	The former hypothesis could point to an intermittent activity scenario, as we discuss in the next sections. 
	However, we cannot exclude that the correlation of \cite{vanvelzen2015} is affected by sample selection effects, since
	different studies have found wider ranges of radio luminosities, or jet powers, for a given accretion-related luminosity, e.g. in the optical \cite[]{mingo2014} or mid-infrared \cite[]{gurkan2014}. Therefore, the estimate $\lbolr$ should not to be overinterpreted \cite[see also][]{hardcastle2018,croston2018}. 
	Still,
it is remarkable that a similar result is obtained, 
with independent methods, both in blazars and in GRGs, i.e. two classes vastly different in physical size and in jet orientation. 
	In the next section we will also show that the estimated jet power is much lower than in radio luminous radio galaxies.

Finally, for the sources having estimates of the black hole mass, we can also estimate the Eddington ratio as the ratio between the bolometric luminosity $\lbolx$ and the Eddington luminosity. In 7 sources out of 8, we estimate Eddington ratios between 0.02 and 0.37, i.e. consistent with the typical values found in efficient HERGs \cite[][]{best2012,mingo2014}. For 4C~34.47 we obtain an Eddington ratio of $\sim 5$, which could indicate that the bolometric luminosity is over-predicted, and/or that the black hole mass is underestimated \cite[\expo{1}{8} \msun\ according to][single-epoch estimate based on the H~$\beta$ line]{liu2006}.
\begin{table*}
	\begin{center}
		\caption{X-ray and radio energetic properties: (2) 2--10 keV flux; (3) 2--10 keV luminosity; (4) core flux density at 1.4 GHz; (5) lobes flux density at 1.4 GHz; (6) core luminosity at 1.4 GHz; (7) lobes luminosity at 1.4 GHz. {\label{tab:radio-x}} }
		\begin{tabular}{l c c c c c c}
			\hline
			Name & $F$(2--10 keV)& $L$(2--10 keV) & $\cflux$ & $\lflux$ & $\clum$ &$\llum$ \\
			&(\tento{-12} \fluxcgs) &(\tento{43} \lumcgs) &(mJy)&(mJy)&(\tento{40} \lumcgs)& (\tento{40} \lumcgs) \\
			\hline
			B3~0309+411b &2.4&13&379&141&25.6&9.5\\[2pt]
			0318+684&3.7&13.5&22&801&0.6&23\\[2pt]
			
			PKS~0707-35(a)&1.2&5&44.6&445.6&2&20.5\\
			PKS~0707-35(b)&&&&1455.4&&67\\[2pt]
			
			4C~73.08 &1.6&2.6&15.6&2521.4&0.18&28.6\\[2pt]
			
			HE~1434-1600&7.2&38&72&7068&5.7&564\\[2pt]
			
			IGR~J14488-4008 & 5.3&25&58.2&276.6&3.3&15.6\\[2pt]
			
			4C~63.22&3.6&45&18.3&657.7&3.1&110\\[2pt]
			
			Mrk 1498&8.1&10&74&557&0.74&5.6\\[2pt]
			
			4C 34.47&8.5&93&610&980&106&171\\[2pt]
			
			IGR~J17488-2338&2.0&42&60.5&337.1&14.9&83.3\\[2pt]
			
			PKS 2014-55&3.9&3.0&40&2560&0.49&31.3\\[2pt]
			
			4C 74.26&3.0&80&184&1621&7.1&62.6\\[2pt]
			
			PKS~2331-240&12.0&6&362&798&2.8&6.1\\[2pt]

			PKS 2356-61 &2.7&11.5&16&24700&0.53&820\\			
			\hline
		\end{tabular}
	\end{center}
\end{table*}

\subsection{Jet power}
\begin{figure}
	\includegraphics[width=\columnwidth]{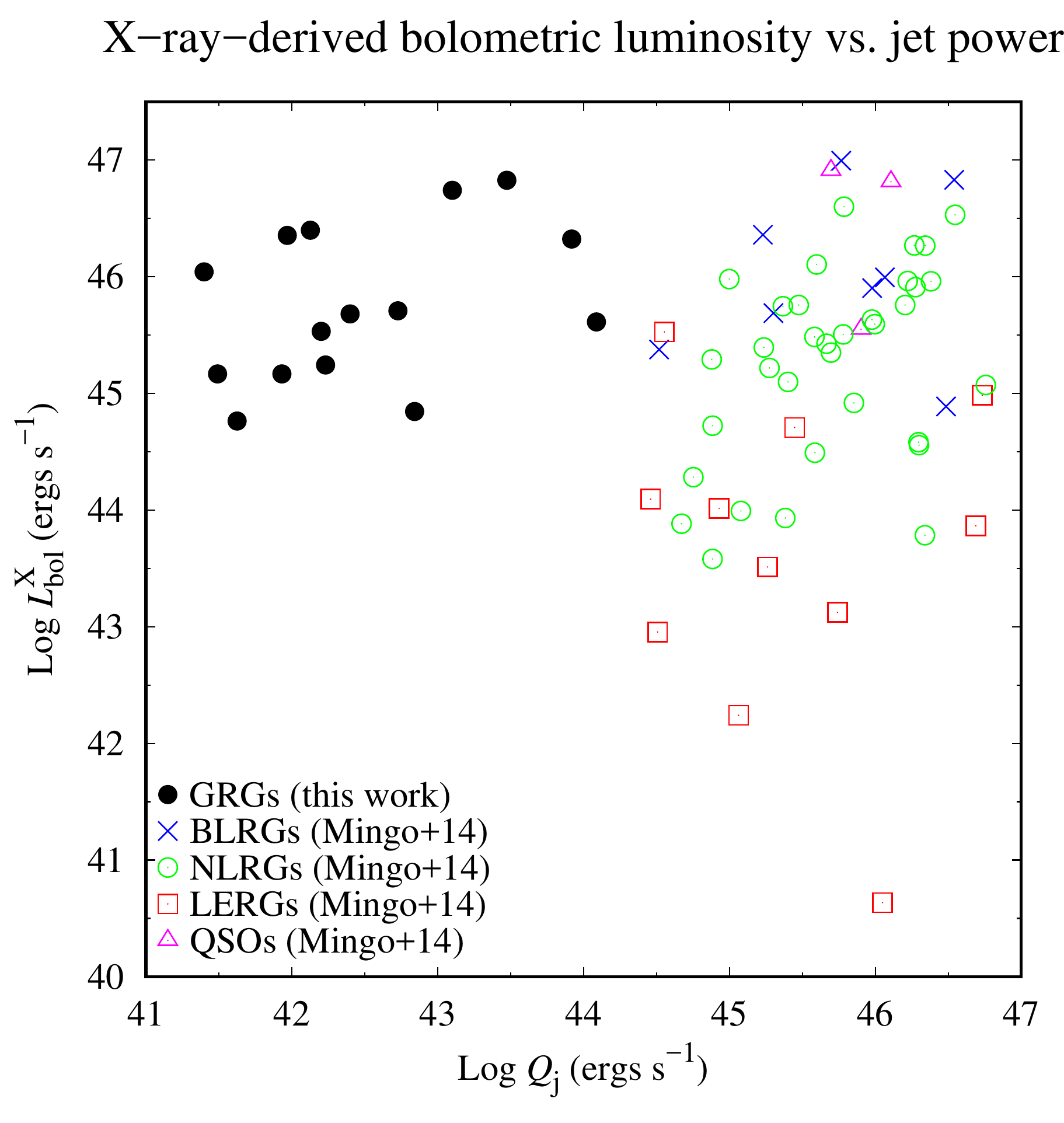}
	\caption{\label{fig:Qj} 
		Bolometric luminosity estimated from the 2--10 keV luminosity versus jet power estimated from the relation of \citet{willott1999}. Black dots denote the GRGs of our sample, overlayed in the plot of \citet{mingo2014}.
		}
\end{figure}

	The time-averaged kinetic power of AGN jets can be estimated from the extended radio luminosity
	\footnote{To this goal, we converted the 1.4-GHz luminosity into the 151-MHz luminosity assuming a spectral index of 0.7.} \cite[]{willott1999,cavagnolo2010}. 
	We estimated the jet power $Q_j$ using the relationship obtained by \cite{willott1999}, based on the minimum energy required in the lobes to produce the observed synchrotron luminosity and the evolution of the radio source.
	In particular, we used
	eq. 12 of \cite{willott1999}:
	\begin{equation}
	Q_j = \expom{3}{38} \mathcal{L}_{151}^{6/7} \textrm{W}
	\end{equation} 
	where $\mathcal{L}_{151}$ is the luminosity at 151 MHz in units of \tento{28} W Hz$^{-1}$ sr$^{-1}$. 
	The uncertainties in the model assumptions are enclosed in a factor $f$, expected to be in the range 1--20, where $Q_j$ depends on $f^{3/2}$. In principle, this would imply a systematic uncertainty up to two orders of magnitude in jet power for a given radio luminosity \cite[see also][]{godfrey2013}. However, the scatter in the radio power/jet power relation is much lower, both from direct measurements \cite[]{godfrey2013} and numerical simulations \cite[at least for sources in a narrow redshift band, see][]{hardcastle2018}. In any case, the relation is useful to probe the efficiency of jet production in different classes of AGNs, as done by \cite{mingo2014}. In Fig. \ref{fig:Qj} we plot the X-ray-derived bolometric luminosity $\lbolx$ against the jet power. 	
	Our GRGs are added on to the plot of the sample of bright radio galaxies of \cite{mingo2014}, selected combining the 2~Jy catalogue \cite[]{2Jy_1} with the 3CRR catalogue \cite[]{laing1983}. We note that \cite{mingo2014}, like us, assumed the \cite{marconi2004} bolometric correction and the \cite{willott1999} relation to derive $\lbolx$ and $Q_j$, respectively.	
	For simplicity, we plot only the sources of the \cite{mingo2014} sample with a well constrained $\lbolx$ (see their Fig. 12). Following \cite{mingo2014}, we highlight different optical classes: broad-line radio galaxies (BLRGs), narrow-line radio galaxies (NLRGs), low-excitation radio galaxies (LERGs) and quasars. 

	From Fig. \ref{fig:Qj}, it can be seen that the X-ray-derived bolometric luminosity of the GRGs of our sample is consistent with that of HERGs (namely BLRGs, NLRGs and quasars) in the \cite{mingo2014} sample. On the other hand, the estimated jet power of our GRGs is found to be down to $\sim \tentom{42}$ \lumcgs, and much lower than that of the sources of \cite{mingo2014}. This is not totally unexpected, because the sources of \cite{mingo2014} are by construction the most radio luminous in the Universe, implying the highest jet powers. However, the difference with our GRGs can be very large, up to 3 orders of magnitude in jet power. Moreover, \cite{mingo2014} also compared their sample with an optically selected sample of radio-loud quasars \cite[]{punsly2011}, and even those sources have an estimated $Q_j > \tentom{43}$ \lumcgs. Therefore, the interpretation of our results is not straightforward.
	First, the (unknown) environment and source age can strongly affect the estimate of the jet power \cite[e.g.][]{shabala2013,hardcastle2018}. Giant sources might have undergone severe radiative losses compared with normal-size radio galaxies, which could explain the relatively small inferred jet power. In other words, any relation between accretion power and ejection at small scales (i.e. the radio core) could be completely lost as the source grows in size, due to radiative losses and complex interaction with the environment. Then, in a luminosity-luminosity diagram like that of Fig. \ref{fig:Qj}, GRGs in the early stage of their life could well start in the upper right part (i.e. high nuclear luminosities/high jet powers) and move to the left side at later times. If in most sources the central engine gradually fades (and eventually switches off), most of radio-selected GRGs would be expected to populate the lower part of the plot (i.e. low nuclear luminosities).
	Hard X-ray selection instead picks the GRGs with a high nuclear luminosity. These could be sources that are able to keep a more or less constant nuclear power during their life (so they just shift horizontally along the diagram), or that experienced one or more episodes of restarting activity. 

\subsection{Restarting activity?}
A number of radio galaxies are known to exhibit signatures of restarting activity, from their morphological and/or spectral radio properties \cite[][and references therein]{SJ2009}.
A striking evidence of restarting activity is seen in the so-called double-double radio galaxies, which exhibit two distinct pairs of radio lobes: a `new' one, closer to the core, and an `old' one, farther away \cite[e.g.][]{schoenmakers_DD}. Among our sample, \pksz\ is a clear example of a restarted radio galaxy \cite[][]{saripalli2013}. PKS~2014-55 is also a restarted giant radio galaxy with an X-shaped morphology that could be a signature of jet reorientation \cite[]{saripalli2007}. An extreme case of jet realignment is \pksa, that switched from a double-lobe GRG to a blazar \cite[]{lorena2017}.
Other hints of restarting jet activity have been found in 4C~73.08, that displays an extended radio cocoon around the brighter lobes \cite[]{wez2016}.
Concerning the spectral properties, some GRGs show a Gigahertz-Peaked Spectrum (GPS) core, which is generally associated with young and compact radio sources at the early stage of their evolution \cite[e.g.][]{odea1998}. This suggests that the core emission is relatively recent, despite the old age of the lobes. Among our sample, this is seen at least in \zero\ \cite[][]{schoenmakers1998} and 4C~74.26 \cite[][]{pearson1992}.  
	A detailed discussion will be presented in Bruni et al. (in prep.), showing that signatures of restarting activity are found in at least 11 sources out of 14, i.e. 80 per cent of the sample.
The restarting activity scenario is an intriguing possibility to explain the properties of hard X-ray selected GRGs. On the one hand, the correlation between radio cores and X-ray emission indicates the existence of a physical relationship between the inner jet at pc scales and the X-ray-emitting region. On the other hand,
	the jet power and the luminosity of the radio lobes
in most sources is less than expected from the nuclear luminosity. Therefore, the sources could be in a restarting phase, characterized by efficient accretion and a 
	high nuclear activity.

Although hard X-ray selection favours the detection of giant objects, no correlation is found between the X-ray luminosity and the source size,
	consistent with the idea 
that more than one parameter is involved in the production of large-scale radio structures \cite[]{bassani2016}. One of such parameters could be the duty cycle.
The intermittency of fuel supply can occur on different time-scales, and has been assumed to explain the optical properties of weak-line radio galaxies \cite[][and references therein]{tadhunter2016}.
If the accretion mechanism in AGNs and black hole X-ray binaries is the same, as suggested by the fundamental plane, then we expect AGNs to undergo dramatic changes of the accretion state like those seen in X-ray binaries \cite[e.g.][]{maccarone2003}.
Given their old age, GRGs offer chances to witness such variations and constrain scenarios of episodic activity. 
Alternatively, if the nuclear activity does not change dramatically over time, the observed dimming of the radio lobes could be due to expansion losses. In particular the inverse Compton scattering of the cosmic microwave background is expected to dominate in giant sources \cite[e.g.][]{ishwara1999,laskar2010,hardcastle2018}. This would lead us to the prediction that the intrinsic fraction of GRGs is much larger than that found in present radio surveys, being greater than the 20 per cent found from hard X-ray selection
\cite[including relict sources not seen in the X-rays and not currently fed by the jet; see][]{saripalli2005}. However, in this case the nuclei are required to stay continuously active for around 100--250 Myr, as indicated by the typical dynamical ages estimated for GRGs \cite[]{machalski2011}.

\section{Summary}\label{sec:summary}
We have presented the first broad-band X-ray study of the nuclei of a sample of 14 hard X-ray selected GRGs, from the literature and the analysis of archival X-ray data. Our main conclusions can be summarized as follows:
\begin{itemize}
	\item[-] The X-ray properties of hard X-ray selected GRGs are analogous to that of normal-size FR II radio galaxies. The X-ray photon index is generally consistent with that of radio-quiet AGNs, and a high-energy cut-off is measured in 4 sources. The bulk of the X-ray emission is generally consistent with originating from a Comptonizing corona, possibly outflowing, coupled to a radiatively efficient accretion flow ($\edd > 0.02$). 
	\item[-] The X-ray luminosity correlates with the radio core luminosity, as expected from the fundamental plane of black hole activity for AGNs and X-ray binaries. However, the relationship is consistent with the `radiatively efficient' branch of the radio--X-ray correlation rather than the `standard/inefficient' branch, consistently with the optical HERG classification. 
	\item[-] 
	The radio luminosity of the lobes is relatively low compared with the nuclear luminosity, as also indicated by the estimate of the jet kinetic power. This can can be explained by restarting activity (i.e. the sources are currently highly accreting and in a high-luminosity state compared with the past activity that produced the old and extended radio lobes) and/or by a significant dimming of the radio lobes due to expansion losses.
\end{itemize}
\section*{Acknowledgements}
We thank the referee for useful comments that improved the paper.
We acknowledge the use of public data from the \nus, \xmm, \swift\ and \integral\ data archives. 
This research has made use of data, software and/or web tools obtained from NASA's High Energy Astrophysics Science Archive Research Center (HEASARC), a service of Goddard Space Flight Center and the Smithsonian Astrophysical Observatory, and of  the \nus\ Data Analysis Software jointly developed by the ASI Space Science Data Center (SSDC, Italy) and the California Institute of
Technology (USA).
We acknowledge financial support from ASI under contracts ASI/INAF 2013-025-R0 and ASI/INAF I/037/12/0 (NARO18).
\bibliographystyle{mnras}
\bibliography{mybib.bib}

\begin{thebibliography}{}
\makeatletter
\relax
\def\mn@urlcharsother{\let\do\@makeother \do\$\do\&\do\#\do\^\do\_\do\%\do\~}
\def\mn@doi{\begingroup\mn@urlcharsother \@ifnextchar [ {\mn@doi@}
  {\mn@doi@[]}}
\def\mn@doi@[#1]#2{\def\@tempa{#1}\ifx\@tempa\@empty \href
  {http://dx.doi.org/#2} {doi:#2}\else \href {http://dx.doi.org/#2} {#1}\fi
  \endgroup}
\def\mn@eprint#1#2{\mn@eprint@#1:#2::\@nil}
\def\mn@eprint@arXiv#1{\href {http://arxiv.org/abs/#1} {{\tt arXiv:#1}}}
\def\mn@eprint@dblp#1{\href {http://dblp.uni-trier.de/rec/bibtex/#1.xml}
  {dblp:#1}}
\def\mn@eprint@#1:#2:#3:#4\@nil{\def\@tempa {#1}\def\@tempb {#2}\def\@tempc
  {#3}\ifx \@tempc \@empty \let \@tempc \@tempb \let \@tempb \@tempa \fi \ifx
  \@tempb \@empty \def\@tempb {arXiv}\fi \@ifundefined
  {mn@eprint@\@tempb}{\@tempb:\@tempc}{\expandafter \expandafter \csname
  mn@eprint@\@tempb\endcsname \expandafter{\@tempc}}}

\bibitem[\protect\citeauthoryear{{Acero} et~al.,}{{Acero}
  et~al.}{2015}]{acero2015}
{Acero} F.,  et~al., 2015, \mn@doi [The Astrophysical Journal Supplement
  Series] {10.1088/0067-0049/218/2/23}, \href
  {https://ui.adsabs.harvard.edu/#abs/2015ApJS..218...23A} {218}

\bibitem[\protect\citeauthoryear{{Anders} \& {Grevesse}}{{Anders} \&
  {Grevesse}}{1989}]{angr}
{Anders} E.,  {Grevesse} N.,  1989, \mn@doi [\gca]
  {10.1016/0016-7037(89)90286-X}, \href
  {http://adsabs.harvard.edu/abs/1989GeCoA..53..197A} {53, 197}

\bibitem[\protect\citeauthoryear{{Antonucci}}{{Antonucci}}{1993}]{antonucci1993}
{Antonucci} R.,  1993, \mn@doi [\araa] {10.1146/annurev.aa.31.090193.002353},
  \href {http://adsabs.harvard.edu/abs/1993ARA%26A..31..473A} {31, 473}

\bibitem[\protect\citeauthoryear{{Arnaud}}{{Arnaud}}{1996}]{arnaud1996}
{Arnaud} K.~A.,  1996, in {Jacoby} G.~H.,  {Barnes} J.,  eds,  Astronomical
  Society of the Pacific Conference Series Vol. 101, Astronomical Data Analysis
  Software and Systems V. p.~17

\bibitem[\protect\citeauthoryear{{Ballantyne} et~al.,}{{Ballantyne}
  et~al.}{2014}]{ballantyne2014}
{Ballantyne} D.~R.,  et~al., 2014, \mn@doi [\apj] {10.1088/0004-637X/794/1/62},
  \href {http://adsabs.harvard.edu/abs/2014ApJ...794...62B} {794, 62}

\bibitem[\protect\citeauthoryear{{Balmaverde}, {Capetti}  \&
  {Grandi}}{{Balmaverde} et~al.}{2006}]{balmaverde2006}
{Balmaverde} B.,  {Capetti} A.,   {Grandi} P.,  2006, \mn@doi [\aap]
  {10.1051/0004-6361:20053799}, \href
  {http://adsabs.harvard.edu/abs/2006A%26A...451...35B} {451, 35}

\bibitem[\protect\citeauthoryear{{Barthelmy} et~al.,}{{Barthelmy}
  et~al.}{2005}]{bat}
{Barthelmy} S.~D.,  et~al., 2005, \mn@doi [\ssr] {10.1007/s11214-005-5096-3},
  \href {http://adsabs.harvard.edu/abs/2005SSRv..120..143B} {120, 143}

\bibitem[\protect\citeauthoryear{{Bassani}, {Venturi}, {Molina}, {Malizia},
  {Dallacasa}, {Panessa}, {Bazzano}  \& {Ubertini}}{{Bassani}
  et~al.}{2016}]{bassani2016}
{Bassani} L.,  {Venturi} T.,  {Molina} M.,  {Malizia} A.,  {Dallacasa} D.,
  {Panessa} F.,  {Bazzano} A.,   {Ubertini} P.,  2016, \mn@doi [\mnras]
  {10.1093/mnras/stw1468}, \href
  {http://adsabs.harvard.edu/abs/2016MNRAS.461.3165B} {461, 3165}

\bibitem[\protect\citeauthoryear{{Baumgartner}, {Tueller}, {Markwardt},
  {Skinner}, {Barthelmy}, {Mushotzky}, {Evans}  \& {Gehrels}}{{Baumgartner}
  et~al.}{2013}]{bat70}
{Baumgartner} W.~H.,  {Tueller} J.,  {Markwardt} C.~B.,  {Skinner} G.~K.,
  {Barthelmy} S.,  {Mushotzky} R.~F.,  {Evans} P.~A.,   {Gehrels} N.,  2013,
  \mn@doi [\apjs] {10.1088/0067-0049/207/2/19}, \href
  {http://adsabs.harvard.edu/abs/2013ApJS..207...19B} {207, 19}

\bibitem[\protect\citeauthoryear{{Beloborodov}}{{Beloborodov}}{1999}]{belo1999}
{Beloborodov} A.~M.,  1999, \mn@doi [\apjl] {10.1086/311810}, \href
  {http://adsabs.harvard.edu/abs/1999ApJ...510L.123B} {510, L123}

\bibitem[\protect\citeauthoryear{{Best} \& {Heckman}}{{Best} \&
  {Heckman}}{2012}]{best2012}
{Best} P.~N.,  {Heckman} T.~M.,  2012, \mn@doi [\mnras]
  {10.1111/j.1365-2966.2012.20414.x}, \href
  {http://adsabs.harvard.edu/abs/2012MNRAS.421.1569B} {421, 1569}

\bibitem[\protect\citeauthoryear{{Bird} et~al.,}{{Bird}
  et~al.}{2010}]{bird2010}
{Bird} A.~J.,  et~al., 2010, \mn@doi [\apjs] {10.1088/0067-0049/186/1/1}, \href
  {http://adsabs.harvard.edu/abs/2010ApJS..186....1B} {186, 1}

\bibitem[\protect\citeauthoryear{{Blandford} \& {Begelman}}{{Blandford} \&
  {Begelman}}{1999}]{adios}
{Blandford} R.~D.,  {Begelman} M.~C.,  1999, \mn@doi [\mnras]
  {10.1046/j.1365-8711.1999.02358.x}, \href
  {http://adsabs.harvard.edu/abs/1999MNRAS.303L...1B} {303, L1}

\bibitem[\protect\citeauthoryear{{Bonchi}, {La Franca}, {Melini}, {Bongiorno}
  \& {Fiore}}{{Bonchi} et~al.}{2013}]{bonchi2013}
{Bonchi} A.,  {La Franca} F.,  {Melini} G.,  {Bongiorno} A.,   {Fiore} F.,
  2013, \mn@doi [\mnras] {10.1093/mnras/sts456}, \href
  {http://adsabs.harvard.edu/abs/2013MNRAS.429.1970B} {429, 1970}

\bibitem[\protect\citeauthoryear{{Cavagnolo}, {McNamara}, {Nulsen}, {Carilli},
  {Jones}  \& {B{\^i}rzan}}{{Cavagnolo} et~al.}{2010}]{cavagnolo2010}
{Cavagnolo} K.~W.,  {McNamara} B.~R.,  {Nulsen} P.~E.~J.,  {Carilli} C.~L.,
  {Jones} C.,   {B{\^i}rzan} L.,  2010, \mn@doi [\apj]
  {10.1088/0004-637X/720/2/1066}, \href
  {http://adsabs.harvard.edu/abs/2010ApJ...720.1066C} {720, 1066}

\bibitem[\protect\citeauthoryear{{Condon}, {Cotton}  \& {Broderick}}{{Condon}
  et~al.}{2002}]{condon2002}
{Condon} J.~J.,  {Cotton} W.~D.,   {Broderick} J.~J.,  2002, \mn@doi [\aj]
  {10.1086/341650}, \href {http://adsabs.harvard.edu/abs/2002AJ....124..675C}
  {124, 675}

\bibitem[\protect\citeauthoryear{{Coriat} et~al.,}{{Coriat}
  et~al.}{2011}]{coriat2011}
{Coriat} M.,  et~al., 2011, \mn@doi [\mnras]
  {10.1111/j.1365-2966.2011.18433.x}, \href
  {http://adsabs.harvard.edu/abs/2011MNRAS.414..677C} {414, 677}

\bibitem[\protect\citeauthoryear{{Croston}, {Ineson}  \&
  {Hardcastle}}{{Croston} et~al.}{2018}]{croston2018}
{Croston} J.~H.,  {Ineson} J.,   {Hardcastle} M.~J.,  2018, \mn@doi [\mnras]
  {10.1093/mnras/sty274}, \href
  {http://adsabs.harvard.edu/abs/2018MNRAS.476.1614C} {476, 1614}

\bibitem[\protect\citeauthoryear{{Di Gesu} \& {Costantini}}{{Di Gesu} \&
  {Costantini}}{2016}]{digesu2016}
{Di Gesu} L.,  {Costantini} E.,  2016, \mn@doi [\aap]
  {10.1051/0004-6361/201628670}, \href
  {http://adsabs.harvard.edu/abs/2016A%26A...594A..88D} {594, A88}

\bibitem[\protect\citeauthoryear{{Eguchi}, {Ueda}, {Terashima}, {Mushotzky}  \&
  {Tueller}}{{Eguchi} et~al.}{2009}]{eguchi2009}
{Eguchi} S.,  {Ueda} Y.,  {Terashima} Y.,  {Mushotzky} R.,   {Tueller} J.,
  2009, \mn@doi [\apj] {10.1088/0004-637X/696/2/1657}, \href
  {http://adsabs.harvard.edu/abs/2009ApJ...696.1657E} {696, 1657}

\bibitem[\protect\citeauthoryear{{Eracleous}, {Sambruna}  \&
  {Mushotzky}}{{Eracleous} et~al.}{2000}]{eracle2000}
{Eracleous} M.,  {Sambruna} R.,   {Mushotzky} R.~F.,  2000, \mn@doi [\apj]
  {10.1086/309076}, \href {http://adsabs.harvard.edu/abs/2000ApJ...537..654E}
  {537, 654}

\bibitem[\protect\citeauthoryear{{Falcke}, {K{\"o}rding}  \&
  {Markoff}}{{Falcke} et~al.}{2004}]{falcke2004}
{Falcke} H.,  {K{\"o}rding} E.,   {Markoff} S.,  2004, \mn@doi [\aap]
  {10.1051/0004-6361:20031683}, \href
  {http://adsabs.harvard.edu/abs/2004A%26A...414..895F} {414, 895}

\bibitem[\protect\citeauthoryear{{Fanaroff} \& {Riley}}{{Fanaroff} \&
  {Riley}}{1974}]{fanaroff&riley}
{Fanaroff} B.~L.,  {Riley} J.~M.,  1974, \mn@doi [\mnras]
  {10.1093/mnras/167.1.31P}, \href
  {http://adsabs.harvard.edu/abs/1974MNRAS.167P..31F} {167, 31P}

\bibitem[\protect\citeauthoryear{{Ferreira}, {Petrucci}, {Henri}, {Saug{\'e}}
  \& {Pelletier}}{{Ferreira} et~al.}{2006}]{ferreira2006}
{Ferreira} J.,  {Petrucci} P.-O.,  {Henri} G.,  {Saug{\'e}} L.,   {Pelletier}
  G.,  2006, \mn@doi [\aap] {10.1051/0004-6361:20052689}, \href
  {http://adsabs.harvard.edu/abs/2006A%26A...447..813F} {447, 813}

\bibitem[\protect\citeauthoryear{{F{\"u}rst} et~al.,}{{F{\"u}rst}
  et~al.}{2016}]{fuerst2016}
{F{\"u}rst} F.,  et~al., 2016, \mn@doi [\apj] {10.3847/0004-637X/819/2/150},
  \href {http://adsabs.harvard.edu/abs/2016ApJ...819..150F} {819, 150}

\bibitem[\protect\citeauthoryear{{Gallo}, {Fender}  \& {Pooley}}{{Gallo}
  et~al.}{2003}]{gallo2003}
{Gallo} E.,  {Fender} R.~P.,   {Pooley} G.~G.,  2003, \mn@doi [\mnras]
  {10.1046/j.1365-8711.2003.06791.x}, \href
  {http://adsabs.harvard.edu/abs/2003MNRAS.344...60G} {344, 60}

\bibitem[\protect\citeauthoryear{{Gallo}, {Miller}  \& {Fender}}{{Gallo}
  et~al.}{2012}]{gallo2012}
{Gallo} E.,  {Miller} B.~P.,   {Fender} R.,  2012, \mn@doi [\mnras]
  {10.1111/j.1365-2966.2012.20899.x}, \href
  {http://adsabs.harvard.edu/abs/2012MNRAS.423..590G} {423, 590}

\bibitem[\protect\citeauthoryear{{George} \& {Fabian}}{{George} \&
  {Fabian}}{1991}]{george&fabian1991}
{George} I.~M.,  {Fabian} A.~C.,  1991, \mnras, \href
  {http://adsabs.harvard.edu/abs/1991MNRAS.249..352G} {249, 352}

\bibitem[\protect\citeauthoryear{{Ghisellini}, {Haardt}  \&
  {Matt}}{{Ghisellini} et~al.}{1994}]{ghm1994}
{Ghisellini} G.,  {Haardt} F.,   {Matt} G.,  1994, \mn@doi [\mnras]
  {10.1093/mnras/267.3.743}, \href
  {http://adsabs.harvard.edu/abs/1994MNRAS.267..743G} {267, 743}

\bibitem[\protect\citeauthoryear{{Ghisellini}, {Tavecchio}, {Maraschi},
  {Celotti}  \& {Sbarrato}}{{Ghisellini} et~al.}{2014}]{ghisellini2014}
{Ghisellini} G.,  {Tavecchio} F.,  {Maraschi} L.,  {Celotti} A.,   {Sbarrato}
  T.,  2014, \mn@doi [\nat] {10.1038/nature13856}, \href
  {http://adsabs.harvard.edu/abs/2014Natur.515..376G} {515, 376}

\bibitem[\protect\citeauthoryear{{Godfrey} \& {Shabala}}{{Godfrey} \&
  {Shabala}}{2013}]{godfrey2013}
{Godfrey} L.~E.~H.,  {Shabala} S.~S.,  2013, \mn@doi [\apj]
  {10.1088/0004-637X/767/1/12}, \href
  {http://adsabs.harvard.edu/abs/2013ApJ...767...12G} {767, 12}

\bibitem[\protect\citeauthoryear{{Grandi}, {Maraschi}, {Urry}  \&
  {Matt}}{{Grandi} et~al.}{2001}]{grandi2001}
{Grandi} P.,  {Maraschi} L.,  {Urry} C.~M.,   {Matt} G.,  2001, \mn@doi [\apj]
  {10.1086/321546}, \href {http://adsabs.harvard.edu/abs/2001ApJ...556...35G}
  {556, 35}

\bibitem[\protect\citeauthoryear{{Grandi}, {Malaguti}  \& {Fiocchi}}{{Grandi}
  et~al.}{2006}]{grandi2006}
{Grandi} P.,  {Malaguti} G.,   {Fiocchi} M.,  2006, \mn@doi [\apj]
  {10.1086/500100}, \href {http://adsabs.harvard.edu/abs/2006ApJ...642..113G}
  {642, 113}

\bibitem[\protect\citeauthoryear{{Guainazzi} \& {Bianchi}}{{Guainazzi} \&
  {Bianchi}}{2007}]{gb2007}
{Guainazzi} M.,  {Bianchi} S.,  2007, \mn@doi [\mnras]
  {10.1111/j.1365-2966.2006.11229.x}, \href
  {http://adsabs.harvard.edu/abs/2007MNRAS.374.1290G} {374, 1290}

\bibitem[\protect\citeauthoryear{{Guainazzi}, {Matt}  \& {Perola}}{{Guainazzi}
  et~al.}{2005}]{gmp2005}
{Guainazzi} M.,  {Matt} G.,   {Perola} G.~C.,  2005, \mn@doi [\aap]
  {10.1051/0004-6361:20053643}, \href
  {http://adsabs.harvard.edu/abs/2005A%26A...444..119G} {444, 119}

\bibitem[\protect\citeauthoryear{{G{\"u}ltekin}, {Cackett}, {Miller}, {Di
  Matteo}, {Markoff}  \& {Richstone}}{{G{\"u}ltekin}
  et~al.}{2009}]{gultekin2009}
{G{\"u}ltekin} K.,  {Cackett} E.~M.,  {Miller} J.~M.,  {Di Matteo} T.,
  {Markoff} S.,   {Richstone} D.~O.,  2009, \mn@doi [\apj]
  {10.1088/0004-637X/706/1/404}, \href
  {http://adsabs.harvard.edu/abs/2009ApJ...706..404G} {706, 404}

\bibitem[\protect\citeauthoryear{{G{\"u}rkan}, {Hardcastle}  \&
  {Jarvis}}{{G{\"u}rkan} et~al.}{2014}]{gurkan2014}
{G{\"u}rkan} G.,  {Hardcastle} M.~J.,   {Jarvis} M.~J.,  2014, \mn@doi [\mnras]
  {10.1093/mnras/stt2264}, \href
  {http://adsabs.harvard.edu/abs/2014MNRAS.438.1149G} {438, 1149}

\bibitem[\protect\citeauthoryear{{Haardt} \& {Maraschi}}{{Haardt} \&
  {Maraschi}}{1991}]{haardt&maraschi1991}
{Haardt} F.,  {Maraschi} L.,  1991, \mn@doi [\apjl] {10.1086/186171}, \href
  {http://adsabs.harvard.edu/abs/1991ApJ...380L..51H} {380, L51}

\bibitem[\protect\citeauthoryear{{Haardt}, {Maraschi}  \&
  {Ghisellini}}{{Haardt} et~al.}{1994}]{hmg1994}
{Haardt} F.,  {Maraschi} L.,   {Ghisellini} G.,  1994, \mn@doi [\apjl]
  {10.1086/187520}, \href {http://adsabs.harvard.edu/abs/1994ApJ...432L..95H}
  {432, L95}

\bibitem[\protect\citeauthoryear{{Haardt}, {Maraschi}  \&
  {Ghisellini}}{{Haardt} et~al.}{1997}]{hmg1997}
{Haardt} F.,  {Maraschi} L.,   {Ghisellini} G.,  1997, \apj, \href
  {http://adsabs.harvard.edu/abs/1997ApJ...476..620H} {476, 620}

\bibitem[\protect\citeauthoryear{{Hardcastle}}{{Hardcastle}}{2018}]{hardcastle2018}
{Hardcastle} M.~J.,  2018, \mn@doi [\mnras] {10.1093/mnras/stx3358}, \href
  {http://adsabs.harvard.edu/abs/2018MNRAS.475.2768H} {475, 2768}

\bibitem[\protect\citeauthoryear{{Hardcastle} \& {Krause}}{{Hardcastle} \&
  {Krause}}{2013}]{hardcastle2013}
{Hardcastle} M.~J.,  {Krause} M.~G.~H.,  2013, \mn@doi [\mnras]
  {10.1093/mnras/sts564}, \href
  {http://adsabs.harvard.edu/abs/2013MNRAS.430..174H} {430, 174}

\bibitem[\protect\citeauthoryear{{Hardcastle}, {Evans}  \&
  {Croston}}{{Hardcastle} et~al.}{2006}]{hardcastle2006}
{Hardcastle} M.~J.,  {Evans} D.~A.,   {Croston} J.~H.,  2006, \mn@doi [\mnras]
  {10.1111/j.1365-2966.2006.10615.x}, \href
  {http://adsabs.harvard.edu/abs/2006MNRAS.370.1893H} {370, 1893}

\bibitem[\protect\citeauthoryear{{Hardcastle}, {Evans}  \&
  {Croston}}{{Hardcastle} et~al.}{2009}]{hardcastle2009}
{Hardcastle} M.~J.,  {Evans} D.~A.,   {Croston} J.~H.,  2009, \mn@doi [\mnras]
  {10.1111/j.1365-2966.2009.14887.x}, \href
  {http://adsabs.harvard.edu/abs/2009MNRAS.396.1929H} {396, 1929}

\bibitem[\protect\citeauthoryear{{Harrison} et~al.,}{{Harrison}
  et~al.}{2013}]{harrison2013nustar}
{Harrison} F.~A.,  et~al., 2013, \mn@doi [\apj] {10.1088/0004-637X/770/2/103},
  \href {http://adsabs.harvard.edu/abs/2013ApJ...770..103H} {770, 103}

\bibitem[\protect\citeauthoryear{{Hern{\'a}ndez-Garc{\'{\i}}a}
  et~al.,}{{Hern{\'a}ndez-Garc{\'{\i}}a} et~al.}{2017}]{lorena2017}
{Hern{\'a}ndez-Garc{\'{\i}}a} L.,  et~al., 2017, \mn@doi [\aap]
  {10.1051/0004-6361/201730530}, \href
  {http://adsabs.harvard.edu/abs/2017A%26A...603A.131H} {603, A131}

\bibitem[\protect\citeauthoryear{{Hern{\'a}ndez-Garc{\'{\i}}a}
  et~al.,}{{Hern{\'a}ndez-Garc{\'{\i}}a} et~al.}{2018}]{lorena2018}
{Hern{\'a}ndez-Garc{\'{\i}}a} L.,  et~al., 2018, \mn@doi [\mnras]
  {10.1093/mnras/sty1345}, \href
  {http://adsabs.harvard.edu/abs/2018MNRAS.478.4634H} {478, 4634}

\bibitem[\protect\citeauthoryear{{Hooper}, {Linden}  \& {Lopez}}{{Hooper}
  et~al.}{2016}]{hooper2016}
{Hooper} D.,  {Linden} T.,   {Lopez} A.,  2016, \mn@doi [\jcap]
  {10.1088/1475-7516/2016/08/019}, \href
  {http://adsabs.harvard.edu/abs/2016JCAP...08..019H} {8, 019}

\bibitem[\protect\citeauthoryear{{Hopkins}, {Richards}  \&
  {Hernquist}}{{Hopkins} et~al.}{2007}]{hopkins2007}
{Hopkins} P.~F.,  {Richards} G.~T.,   {Hernquist} L.,  2007, \mn@doi [\apj]
  {10.1086/509629}, \href {http://adsabs.harvard.edu/abs/2007ApJ...654..731H}
  {654, 731}

\bibitem[\protect\citeauthoryear{{Ishwara-Chandra} \&
  {Saikia}}{{Ishwara-Chandra} \& {Saikia}}{1999}]{ishwara1999}
{Ishwara-Chandra} C.~H.,  {Saikia} D.~J.,  1999, \mn@doi [\mnras]
  {10.1046/j.1365-8711.1999.02835.x}, \href
  {http://adsabs.harvard.edu/abs/1999MNRAS.309..100I} {309, 100}

\bibitem[\protect\citeauthoryear{{Jagers}, {Miley}, {van Breugel}, {Schilizzi}
  \& {Conway}}{{Jagers} et~al.}{1982}]{jagers1982}
{Jagers} W.~J.,  {Miley} G.~K.,  {van Breugel} W.~J.~M.,  {Schilizzi} R.~T.,
  {Conway} R.~G.,  1982, \aap, \href
  {http://adsabs.harvard.edu/abs/1982A%26A...105..278J} {105, 278}

\bibitem[\protect\citeauthoryear{{Jansen} et~al.,}{{Jansen} et~al.}{2001}]{xmm}
{Jansen} F.,  et~al., 2001, \mn@doi [\aap] {10.1051/0004-6361:20000036}, \href
  {http://adsabs.harvard.edu/abs/2001A%26A...365L...1J} {365, L1}

\bibitem[\protect\citeauthoryear{{Kaiser} \& {Alexander}}{{Kaiser} \&
  {Alexander}}{1999}]{kaiser1999}
{Kaiser} C.~R.,  {Alexander} P.,  1999, \mn@doi [\mnras]
  {10.1046/j.1365-8711.1999.02129.x}, \href
  {http://adsabs.harvard.edu/abs/1999MNRAS.302..515K} {302, 515}

\bibitem[\protect\citeauthoryear{{Kalberla}, {Burton}, {Hartmann}, {Arnal},
  {Bajaja}, {Morras}  \& {P{\"o}ppel}}{{Kalberla} et~al.}{2005}]{kalberla2005}
{Kalberla} P.~M.~W.,  {Burton} W.~B.,  {Hartmann} D.,  {Arnal} E.~M.,  {Bajaja}
  E.,  {Morras} R.,   {P{\"o}ppel} W.~G.~L.,  2005, \mn@doi [\aap]
  {10.1051/0004-6361:20041864}, \href
  {http://adsabs.harvard.edu/abs/2005A%26A...440..775K} {440, 775}

\bibitem[\protect\citeauthoryear{{Keel} et~al.,}{{Keel}
  et~al.}{2015}]{keel2015}
{Keel} W.~C.,  et~al., 2015, \mn@doi [\aj] {10.1088/0004-6256/149/5/155}, \href
  {http://adsabs.harvard.edu/abs/2015AJ....149..155K} {149, 155}

\bibitem[\protect\citeauthoryear{{Keel} et~al.,}{{Keel}
  et~al.}{2017}]{keel2017}
{Keel} W.~C.,  et~al., 2017, \mn@doi [\apj] {10.3847/1538-4357/835/2/256},
  \href {http://adsabs.harvard.edu/abs/2017ApJ...835..256K} {835, 256}

\bibitem[\protect\citeauthoryear{{King}, {Lohfink}  \& {Kara}}{{King}
  et~al.}{2017}]{king2017}
{King} A.~L.,  {Lohfink} A.,   {Kara} E.,  2017, \mn@doi [\apj]
  {10.3847/1538-4357/835/2/226}, \href
  {http://adsabs.harvard.edu/abs/2017ApJ...835..226K} {835, 226}

\bibitem[\protect\citeauthoryear{{K{\"o}rding}, {Falcke}  \&
  {Corbel}}{{K{\"o}rding} et~al.}{2006}]{kording2006}
{K{\"o}rding} E.,  {Falcke} H.,   {Corbel} S.,  2006, \mn@doi [\aap]
  {10.1051/0004-6361:20054144}, \href
  {http://adsabs.harvard.edu/abs/2006A%26A...456..439K} {456, 439}

\bibitem[\protect\citeauthoryear{{K{\"o}rding}, {Jester}  \&
  {Fender}}{{K{\"o}rding} et~al.}{2008}]{kording2008}
{K{\"o}rding} E.~G.,  {Jester} S.,   {Fender} R.,  2008, \mn@doi [\mnras]
  {10.1111/j.1365-2966.2007.12529.x}, \href
  {http://adsabs.harvard.edu/abs/2008MNRAS.383..277K} {383, 277}

\bibitem[\protect\citeauthoryear{{Laing}, {Riley}  \& {Longair}}{{Laing}
  et~al.}{1983}]{laing1983}
{Laing} R.~A.,  {Riley} J.~M.,   {Longair} M.~S.,  1983, \mn@doi [\mnras]
  {10.1093/mnras/204.1.151}, \href
  {http://adsabs.harvard.edu/abs/1983MNRAS.204..151L} {204, 151}

\bibitem[\protect\citeauthoryear{{Lara}, {Cotton}, {Feretti}, {Giovannini},
  {Marcaide}, {M{\'a}rquez}  \& {Venturi}}{{Lara} et~al.}{2001}]{lara2001}
{Lara} L.,  {Cotton} W.~D.,  {Feretti} L.,  {Giovannini} G.,  {Marcaide} J.~M.,
   {M{\'a}rquez} I.,   {Venturi} T.,  2001, \mn@doi [\aap]
  {10.1051/0004-6361:20010254}, \href
  {http://adsabs.harvard.edu/abs/2001A%26A...370..409L} {370, 409}

\bibitem[\protect\citeauthoryear{{Laskar}, {Fabian}, {Blundell}  \&
  {Erlund}}{{Laskar} et~al.}{2010}]{laskar2010}
{Laskar} T.,  {Fabian} A.~C.,  {Blundell} K.~M.,   {Erlund} M.~C.,  2010,
  \mn@doi [\mnras] {10.1111/j.1365-2966.2009.15769.x}, \href
  {http://adsabs.harvard.edu/abs/2010MNRAS.401.1500L} {401, 1500}

\bibitem[\protect\citeauthoryear{{Letawe}, {Courbin}, {Magain}, {Hilker},
  {Jablonka}, {Jahnke}  \& {Wisotzki}}{{Letawe} et~al.}{2004}]{letawe2004}
{Letawe} G.,  {Courbin} F.,  {Magain} P.,  {Hilker} M.,  {Jablonka} P.,
  {Jahnke} K.,   {Wisotzki} L.,  2004, \mn@doi [\aap]
  {10.1051/0004-6361:20040481}, \href
  {http://adsabs.harvard.edu/abs/2004A%26A...424..455L} {424, 455}

\bibitem[\protect\citeauthoryear{{Li}, {Wu}  \& {Wang}}{{Li}
  et~al.}{2008}]{li2008}
{Li} Z.-Y.,  {Wu} X.-B.,   {Wang} R.,  2008, \mn@doi [\apj] {10.1086/592314},
  \href {http://adsabs.harvard.edu/abs/2008ApJ...688..826L} {688, 826}

\bibitem[\protect\citeauthoryear{{Liu}, {Jiang}  \& {Gu}}{{Liu}
  et~al.}{2006}]{liu2006}
{Liu} Y.,  {Jiang} D.~R.,   {Gu} M.~F.,  2006, \mn@doi [\apj] {10.1086/498639},
  \href {http://adsabs.harvard.edu/abs/2006ApJ...637..669L} {637, 669}

\bibitem[\protect\citeauthoryear{{Lohfink} et~al.,}{{Lohfink}
  et~al.}{2013}]{lohfink2013}
{Lohfink} A.~M.,  et~al., 2013, \mn@doi [\apj] {10.1088/0004-637X/772/2/83},
  \href {http://adsabs.harvard.edu/abs/2013ApJ...772...83L} {772, 83}

\bibitem[\protect\citeauthoryear{{Lohfink} et~al.,}{{Lohfink}
  et~al.}{2015}]{lohfink2015}
{Lohfink} A.~M.,  et~al., 2015, \mn@doi [\apj] {10.1088/0004-637X/814/1/24},
  \href {http://adsabs.harvard.edu/abs/2015ApJ...814...24L} {814, 24}

\bibitem[\protect\citeauthoryear{{Lohfink} et~al.,}{{Lohfink}
  et~al.}{2017}]{lohfink2017}
{Lohfink} A.~M.,  et~al., 2017, \mn@doi [\apj] {10.3847/1538-4357/aa6d07},
  \href {http://adsabs.harvard.edu/abs/2017ApJ...841...80L} {841, 80}

\bibitem[\protect\citeauthoryear{{Maccarone}, {Gallo}  \& {Fender}}{{Maccarone}
  et~al.}{2003}]{maccarone2003}
{Maccarone} T.~J.,  {Gallo} E.,   {Fender} R.,  2003, \mn@doi [\mnras]
  {10.1046/j.1365-8711.2003.07161.x}, \href
  {http://adsabs.harvard.edu/abs/2003MNRAS.345L..19M} {345, L19}

\bibitem[\protect\citeauthoryear{{Machalski}}{{Machalski}}{2011}]{machalski2011}
{Machalski} J.,  2011, \mn@doi [\mnras] {10.1111/j.1365-2966.2011.18314.x},
  \href {http://adsabs.harvard.edu/abs/2011MNRAS.413.2429M} {413, 2429}

\bibitem[\protect\citeauthoryear{{Machalski}, {Chyzy}  \&
  {Jamrozy}}{{Machalski} et~al.}{2004}]{machalski2004}
{Machalski} J.,  {Chyzy} K.~T.,   {Jamrozy} M.,  2004, \actaa, \href
  {http://adsabs.harvard.edu/abs/2004AcA....54..249M} {54, 249}

\bibitem[\protect\citeauthoryear{{Mack}, {Klein}, {O'Dea}, {Willis}  \&
  {Saripalli}}{{Mack} et~al.}{1998}]{mack1998}
{Mack} K.-H.,  {Klein} U.,  {O'Dea} C.~P.,  {Willis} A.~G.,   {Saripalli} L.,
  1998, \aap, \href {http://adsabs.harvard.edu/abs/1998A%26A...329..431M} {329,
  431}

\bibitem[\protect\citeauthoryear{{Madsen} et~al.,}{{Madsen}
  et~al.}{2015}]{madsen2015}
{Madsen} K.~K.,  et~al., 2015, \mn@doi [\apj] {10.1088/0004-637X/812/1/14},
  \href {http://adsabs.harvard.edu/abs/2015ApJ...812...14M} {812, 14}

\bibitem[\protect\citeauthoryear{{Magdziarz} \& {Zdziarski}}{{Magdziarz} \&
  {Zdziarski}}{1995}]{pexrav}
{Magdziarz} P.,  {Zdziarski} A.~A.,  1995, \mnras, \href
  {http://adsabs.harvard.edu/abs/1995MNRAS.273..837M} {273, 837}

\bibitem[\protect\citeauthoryear{{Malarecki}, {Staveley-Smith}, {Saripalli},
  {Subrahmanyan}, {Jones}, {Duffy}  \& {Rioja}}{{Malarecki}
  et~al.}{2013}]{malarecki2013}
{Malarecki} J.~M.,  {Staveley-Smith} L.,  {Saripalli} L.,  {Subrahmanyan} R.,
  {Jones} D.~H.,  {Duffy} A.~R.,   {Rioja} M.,  2013, \mn@doi [\mnras]
  {10.1093/mnras/stt471}, \href
  {http://adsabs.harvard.edu/abs/2013MNRAS.432..200M} {432, 200}

\bibitem[\protect\citeauthoryear{{Malarecki}, {Jones}, {Saripalli},
  {Staveley-Smith}  \& {Subrahmanyan}}{{Malarecki}
  et~al.}{2015}]{malarecki2015}
{Malarecki} J.~M.,  {Jones} D.~H.,  {Saripalli} L.,  {Staveley-Smith} L.,
  {Subrahmanyan} R.,  2015, \mn@doi [\mnras] {10.1093/mnras/stv273}, \href
  {http://adsabs.harvard.edu/abs/2015MNRAS.449..955M} {449, 955}

\bibitem[\protect\citeauthoryear{{Malizia}, {Bassani}, {Bazzano}, {Bird},
  {Masetti}, {Panessa}, {Stephen}  \& {Ubertini}}{{Malizia}
  et~al.}{2012}]{malizia2012}
{Malizia} A.,  {Bassani} L.,  {Bazzano} A.,  {Bird} A.~J.,  {Masetti} N.,
  {Panessa} F.,  {Stephen} J.~B.,   {Ubertini} P.,  2012, \mn@doi [\mnras]
  {10.1111/j.1365-2966.2012.21755.x}, \href
  {http://adsabs.harvard.edu/abs/2012MNRAS.426.1750M} {426, 1750}

\bibitem[\protect\citeauthoryear{Malizia, Molina, Bassani, Stephen, Bazzano,
  Ubertini  \& Bird}{Malizia et~al.}{2014}]{malizia2014}
Malizia A.,  Molina M.,  Bassani L.,  Stephen J.~B.,  Bazzano A.,  Ubertini P.,
    Bird A.~J.,  2014, \apj, 782, L25

\bibitem[\protect\citeauthoryear{{Malzac}, {Beloborodov}  \&
  {Poutanen}}{{Malzac} et~al.}{2001}]{malzac2001}
{Malzac} J.,  {Beloborodov} A.~M.,   {Poutanen} J.,  2001, \mn@doi [\mnras]
  {10.1046/j.1365-8711.2001.04450.x}, \href
  {http://adsabs.harvard.edu/abs/2001MNRAS.326..417M} {326, 417}

\bibitem[\protect\citeauthoryear{{Marconi}, {Risaliti}, {Gilli}, {Hunt},
  {Maiolino}  \& {Salvati}}{{Marconi} et~al.}{2004}]{marconi2004}
{Marconi} A.,  {Risaliti} G.,  {Gilli} R.,  {Hunt} L.~K.,  {Maiolino} R.,
  {Salvati} M.,  2004, \mn@doi [\mnras] {10.1111/j.1365-2966.2004.07765.x},
  \href {http://adsabs.harvard.edu/abs/2004MNRAS.351..169M} {351, 169}

\bibitem[\protect\citeauthoryear{{Markoff}, {Nowak}  \& {Wilms}}{{Markoff}
  et~al.}{2005}]{markoff2005}
{Markoff} S.,  {Nowak} M.~A.,   {Wilms} J.,  2005, \mn@doi [\apj]
  {10.1086/497628}, \href {http://adsabs.harvard.edu/abs/2005ApJ...635.1203M}
  {635, 1203}

\bibitem[\protect\citeauthoryear{{Matt}, {Perola}  \& {Piro}}{{Matt}
  et~al.}{1991}]{MPP1991}
{Matt} G.,  {Perola} G.~C.,   {Piro} L.,  1991, \aap, \href
  {http://adsabs.harvard.edu/abs/1991A%26A...247...25M} {247, 25}

\bibitem[\protect\citeauthoryear{{Matt}, {Guainazzi}  \& {Maiolino}}{{Matt}
  et~al.}{2003}]{matt2003}
{Matt} G.,  {Guainazzi} M.,   {Maiolino} R.,  2003, \mn@doi [\mnras]
  {10.1046/j.1365-8711.2003.06539.x}, \href
  {http://adsabs.harvard.edu/abs/2003MNRAS.342..422M} {342, 422}

\bibitem[\protect\citeauthoryear{{Merloni}, {Heinz}  \& {di Matteo}}{{Merloni}
  et~al.}{2003}]{merloni2003}
{Merloni} A.,  {Heinz} S.,   {di Matteo} T.,  2003, \mn@doi [\mnras]
  {10.1046/j.1365-2966.2003.07017.x}, \href
  {http://adsabs.harvard.edu/abs/2003MNRAS.345.1057M} {345, 1057}

\bibitem[\protect\citeauthoryear{{Merloni}, {K{\"o}rding}, {Heinz}, {Markoff},
  {Di Matteo}  \& {Falcke}}{{Merloni} et~al.}{2006}]{merloni2006}
{Merloni} A.,  {K{\"o}rding} E.,  {Heinz} S.,  {Markoff} S.,  {Di Matteo} T.,
  {Falcke} H.,  2006, \mn@doi [\na] {10.1016/j.newast.2006.03.002}, \href
  {http://adsabs.harvard.edu/abs/2006NewA...11..567M} {11, 567}

\bibitem[\protect\citeauthoryear{{Mingo}, {Hardcastle}, {Croston}, {Dicken},
  {Evans}, {Morganti}  \& {Tadhunter}}{{Mingo} et~al.}{2014}]{mingo2014}
{Mingo} B.,  {Hardcastle} M.~J.,  {Croston} J.~H.,  {Dicken} D.,  {Evans}
  D.~A.,  {Morganti} R.,   {Tadhunter} C.,  2014, \mn@doi [\mnras]
  {10.1093/mnras/stu263}, \href
  {http://adsabs.harvard.edu/abs/2014MNRAS.440..269M} {440, 269}

\bibitem[\protect\citeauthoryear{{Mingo} et~al.,}{{Mingo}
  et~al.}{2017}]{mingo2017}
{Mingo} B.,  et~al., 2017, \mn@doi [\mnras] {10.1093/mnras/stx1307}, \href
  {http://adsabs.harvard.edu/abs/2017MNRAS.470.2762M} {470, 2762}

\bibitem[\protect\citeauthoryear{{Molina} et~al.,}{{Molina}
  et~al.}{2008}]{molina2008}
{Molina} M.,  et~al., 2008, \mn@doi [\mnras]
  {10.1111/j.1365-2966.2008.13824.x}, \href
  {http://adsabs.harvard.edu/abs/2008MNRAS.390.1217M} {390, 1217}

\bibitem[\protect\citeauthoryear{{Molina}, {Bassani}, {Malizia}, {Stephen},
  {Bird}, {Bazzano}  \& {Ubertini}}{{Molina} et~al.}{2013}]{molina2013}
{Molina} M.,  {Bassani} L.,  {Malizia} A.,  {Stephen} J.~B.,  {Bird} A.~J.,
  {Bazzano} A.,   {Ubertini} P.,  2013, \mn@doi [\mnras]
  {10.1093/mnras/stt844}, \href
  {http://adsabs.harvard.edu/abs/2013MNRAS.433.1687M} {433, 1687}

\bibitem[\protect\citeauthoryear{{Molina}, {Bassani}, {Malizia}, {Bird},
  {Bazzano}, {Ubertini}  \& {Venturi}}{{Molina} et~al.}{2014}]{molina_17488}
{Molina} M.,  {Bassani} L.,  {Malizia} A.,  {Bird} A.~J.,  {Bazzano} A.,
  {Ubertini} P.,   {Venturi} T.,  2014, \mn@doi [\aap]
  {10.1051/0004-6361/201423609}, \href
  {http://adsabs.harvard.edu/abs/2014A%26A...565A...2M} {565, A2}

\bibitem[\protect\citeauthoryear{{Molina}, {Venturi}, {Malizia}, {Bassani},
  {Dallacasa}, {Lal}, {Bird}  \& {Ubertini}}{{Molina}
  et~al.}{2015}]{molina_14488}
{Molina} M.,  {Venturi} T.,  {Malizia} A.,  {Bassani} L.,  {Dallacasa} D.,
  {Lal} D.~V.,  {Bird} A.~J.,   {Ubertini} P.,  2015, \mn@doi [\mnras]
  {10.1093/mnras/stv1116}, \href
  {http://adsabs.harvard.edu/abs/2015MNRAS.451.2370M} {451, 2370}

\bibitem[\protect\citeauthoryear{{Morganti}, {Killeen}  \&
  {Tadhunter}}{{Morganti} et~al.}{1993}]{morganti1993}
{Morganti} R.,  {Killeen} N.~E.~B.,   {Tadhunter} C.~N.,  1993, \mn@doi
  [\mnras] {10.1093/mnras/263.4.1023}, \href
  {http://adsabs.harvard.edu/abs/1993MNRAS.263.1023M} {263, 1023}

\bibitem[\protect\citeauthoryear{{Morganti}, {Oosterloo}, {Reynolds},
  {Tadhunter}  \& {Migenes}}{{Morganti} et~al.}{1997}]{morganti1997}
{Morganti} R.,  {Oosterloo} T.~A.,  {Reynolds} J.~E.,  {Tadhunter} C.~N.,
  {Migenes} V.,  1997, \mn@doi [\mnras] {10.1093/mnras/284.3.541}, \href
  {http://adsabs.harvard.edu/abs/1997MNRAS.284..541M} {284, 541}

\bibitem[\protect\citeauthoryear{{Motta}, {Casella}  \& {Fender}}{{Motta}
  et~al.}{2018}]{motta2018}
{Motta} S.~E.,  {Casella} P.,   {Fender} R.~P.,  2018, \mn@doi [\mnras]
  {10.1093/mnras/sty1440}, \href
  {http://adsabs.harvard.edu/abs/2018MNRAS.478.5159M} {478, 5159}

\bibitem[\protect\citeauthoryear{{Mushotzy}, {Winter}, {McIntosh}  \&
  {Tueller}}{{Mushotzy} et~al.}{2008}]{mushotzky2008}
{Mushotzy} R.~F.,  {Winter} L.~M.,  {McIntosh} D.~H.,   {Tueller} J.,  2008,
  \mn@doi [\apjl] {10.1086/592196}, \href
  {http://adsabs.harvard.edu/abs/2008ApJ...684L..65M} {684, L65}

\bibitem[\protect\citeauthoryear{{Narayan} \& {Yi}}{{Narayan} \&
  {Yi}}{1994}]{adaf}
{Narayan} R.,  {Yi} I.,  1994, \mn@doi [\apjl] {10.1086/187381}, \href
  {http://adsabs.harvard.edu/abs/1994ApJ...428L..13N} {428, L13}

\bibitem[\protect\citeauthoryear{{O'Dea}}{{O'Dea}}{1998}]{odea1998}
{O'Dea} C.~P.,  1998, \mn@doi [\pasp] {10.1086/316162}, \href
  {http://adsabs.harvard.edu/abs/1998PASP..110..493O} {110, 493}

\bibitem[\protect\citeauthoryear{{Oh} et~al.,}{{Oh} et~al.}{2018}]{bat105}
{Oh} K.,  et~al., 2018, \mn@doi [\apjs] {10.3847/1538-4365/aaa7fd}, \href
  {http://adsabs.harvard.edu/abs/2018ApJS..235....4O} {235, 4}

\bibitem[\protect\citeauthoryear{{Panessa} et~al.,}{{Panessa}
  et~al.}{2015}]{panessa2015}
{Panessa} F.,  et~al., 2015, \mn@doi [\mnras] {10.1093/mnras/stu2455}, \href
  {http://adsabs.harvard.edu/abs/2015MNRAS.447.1289P} {447, 1289}

\bibitem[\protect\citeauthoryear{{Panessa} et~al.,}{{Panessa}
  et~al.}{2016}]{panessa2016}
{Panessa} F.,  et~al., 2016, \mn@doi [\mnras] {10.1093/mnras/stw1438}, \href
  {http://adsabs.harvard.edu/abs/2016MNRAS.461.3153P} {461, 3153}

\bibitem[\protect\citeauthoryear{{Parisi} et~al.,}{{Parisi}
  et~al.}{2012}]{parisi2012}
{Parisi} P.,  et~al., 2012, \mn@doi [\aap] {10.1051/0004-6361/201219192}, \href
  {http://adsabs.harvard.edu/abs/2012A%26A...545A.101P} {545, A101}

\bibitem[\protect\citeauthoryear{{Pearson}, {Blundell}, {Riley}  \&
  {Warner}}{{Pearson} et~al.}{1992}]{pearson1992}
{Pearson} T.~J.,  {Blundell} K.~M.,  {Riley} J.~M.,   {Warner} P.~J.,  1992,
  \mn@doi [\mnras] {10.1093/mnras/259.1.13P}, \href
  {http://adsabs.harvard.edu/abs/1992MNRAS.259P..13P} {259, 13P}

\bibitem[\protect\citeauthoryear{{Petrucci}, {Ferreira}, {Henri}, {Malzac}  \&
  {Foellmi}}{{Petrucci} et~al.}{2010}]{pop2010}
{Petrucci} P.~O.,  {Ferreira} J.,  {Henri} G.,  {Malzac} J.,   {Foellmi} C.,
  2010, \mn@doi [\aap] {10.1051/0004-6361/201014753}, \href
  {http://adsabs.harvard.edu/abs/2010A%26A...522A..38P} {522, A38}

\bibitem[\protect\citeauthoryear{{Piconcelli}, {Jimenez-Bail{\'o}n},
  {Guainazzi}, {Schartel}, {Rodr{\'{\i}}guez-Pascual}  \&
  {Santos-Lle{\'o}}}{{Piconcelli} et~al.}{2004}]{pico2004}
{Piconcelli} E.,  {Jimenez-Bail{\'o}n} E.,  {Guainazzi} M.,  {Schartel} N.,
  {Rodr{\'{\i}}guez-Pascual} P.~M.,   {Santos-Lle{\'o}} M.,  2004, \mn@doi
  [\mnras] {10.1111/j.1365-2966.2004.07764.x}, \href
  {http://adsabs.harvard.edu/abs/2004MNRAS.351..161P} {351, 161}

\bibitem[\protect\citeauthoryear{{Pjanka}, {Zdziarski}  \& {Sikora}}{{Pjanka}
  et~al.}{2017}]{pjanka2017}
{Pjanka} P.,  {Zdziarski} A.~A.,   {Sikora} M.,  2017, \mn@doi [\mnras]
  {10.1093/mnras/stw2960}, \href
  {http://adsabs.harvard.edu/abs/2017MNRAS.465.3506P} {465, 3506}

\bibitem[\protect\citeauthoryear{{Punsly} \& {Zhang}}{{Punsly} \&
  {Zhang}}{2011}]{punsly2011}
{Punsly} B.,  {Zhang} S.,  2011, \mn@doi [\apjl] {10.1088/2041-8205/735/1/L3},
  \href {http://adsabs.harvard.edu/abs/2011ApJ...735L...3P} {735, L3}

\bibitem[\protect\citeauthoryear{{Reynolds} et~al.,}{{Reynolds}
  et~al.}{2015}]{reynolds2015}
{Reynolds} C.~S.,  et~al., 2015, \mn@doi [\apj] {10.1088/0004-637X/808/2/154},
  \href {http://adsabs.harvard.edu/abs/2015ApJ...808..154R} {808, 154}

\bibitem[\protect\citeauthoryear{{Ricci} et~al.,}{{Ricci}
  et~al.}{2017}]{ricci2017}
{Ricci} C.,  et~al., 2017, \mn@doi [\apjs] {10.3847/1538-4365/aa96ad}, \href
  {http://adsabs.harvard.edu/abs/2017ApJS..233...17R} {233, 17}

\bibitem[\protect\citeauthoryear{{Runnoe}, {Brotherton}  \& {Shang}}{{Runnoe}
  et~al.}{2012}]{runnoe2012}
{Runnoe} J.~C.,  {Brotherton} M.~S.,   {Shang} Z.,  2012, \mn@doi [\mnras]
  {10.1111/j.1365-2966.2012.20620.x}, \href
  {http://adsabs.harvard.edu/abs/2012MNRAS.422..478R} {422, 478}

\bibitem[\protect\citeauthoryear{{Saikia} \& {Jamrozy}}{{Saikia} \&
  {Jamrozy}}{2009}]{SJ2009}
{Saikia} D.~J.,  {Jamrozy} M.,  2009, Bulletin of the Astronomical Society of
  India, \href {http://adsabs.harvard.edu/abs/2009BASI...37...63S} {37}

\bibitem[\protect\citeauthoryear{{Saripalli} \& {Subrahmanyan}}{{Saripalli} \&
  {Subrahmanyan}}{2009}]{saripalli2009}
{Saripalli} L.,  {Subrahmanyan} R.,  2009, \mn@doi [\apj]
  {10.1088/0004-637X/695/1/156}, \href
  {http://adsabs.harvard.edu/abs/2009ApJ...695..156S} {695, 156}

\bibitem[\protect\citeauthoryear{{Saripalli}, {Hunstead}, {Subrahmanyan}  \&
  {Boyce}}{{Saripalli} et~al.}{2005}]{saripalli2005}
{Saripalli} L.,  {Hunstead} R.~W.,  {Subrahmanyan} R.,   {Boyce} E.,  2005,
  \mn@doi [\aj] {10.1086/432507}, \href
  {http://adsabs.harvard.edu/abs/2005AJ....130..896S} {130, 896}

\bibitem[\protect\citeauthoryear{{Saripalli}, {Subrahmanyan}, {Laskar}  \&
  {Koekemoer}}{{Saripalli} et~al.}{2007}]{saripalli2007}
{Saripalli} L.,  {Subrahmanyan} R.,  {Laskar} T.,   {Koekemoer} A.,  2007, in
  Proc. Science (MRU), 052, From Planets to Dark Energy: the Modern Radio
  Universe. p.~130

\bibitem[\protect\citeauthoryear{{Saripalli}, {Subrahmanyan}, {Thorat},
  {Ekers}, {Hunstead}, {Johnston}  \& {Sadler}}{{Saripalli}
  et~al.}{2012}]{saripalli2012}
{Saripalli} L.,  {Subrahmanyan} R.,  {Thorat} K.,  {Ekers} R.~D.,  {Hunstead}
  R.~W.,  {Johnston} H.~M.,   {Sadler} E.~M.,  2012, \mn@doi [\apjs]
  {10.1088/0067-0049/199/2/27}, \href
  {http://adsabs.harvard.edu/abs/2012ApJS..199...27S} {199, 27}

\bibitem[\protect\citeauthoryear{{Saripalli}, {Malarecki}, {Subrahmanyan},
  {Jones}  \& {Staveley-Smith}}{{Saripalli} et~al.}{2013}]{saripalli2013}
{Saripalli} L.,  {Malarecki} J.~M.,  {Subrahmanyan} R.,  {Jones} D.~H.,
  {Staveley-Smith} L.,  2013, \mn@doi [\mnras] {10.1093/mnras/stt1606}, \href
  {http://adsabs.harvard.edu/abs/2013MNRAS.436..690S} {436, 690}

\bibitem[\protect\citeauthoryear{{Schoenmakers}, {Mack}, {Lara},
  {R{\"o}ttgering}, {de Bruyn}, {van der Laan}  \& {Giovannini}}{{Schoenmakers}
  et~al.}{1998}]{schoenmakers1998}
{Schoenmakers} A.~P.,  {Mack} K.-H.,  {Lara} L.,  {R{\"o}ttgering} H.~J.~A.,
  {de Bruyn} A.~G.,  {van der Laan} H.,   {Giovannini} G.,  1998, \aap, \href
  {http://adsabs.harvard.edu/abs/1998A%26A...336..455S} {336, 455}

\bibitem[\protect\citeauthoryear{{Schoenmakers}, {Mack}, {de Bruyn},
  {R{\"o}ttgering}, {Klein}  \& {van der Laan}}{{Schoenmakers}
  et~al.}{2000a}]{schoenmakers2000}
{Schoenmakers} A.~P.,  {Mack} K.-H.,  {de Bruyn} A.~G.,  {R{\"o}ttgering}
  H.~J.~A.,  {Klein} U.,   {van der Laan} H.,  2000a, \mn@doi [\aaps]
  {10.1051/aas:2000267}, \href
  {http://adsabs.harvard.edu/abs/2000A%26AS..146..293S} {146, 293}

\bibitem[\protect\citeauthoryear{{Schoenmakers}, {de Bruyn}, {R{\"o}ttgering},
  {van der Laan}  \& {Kaiser}}{{Schoenmakers} et~al.}{2000b}]{schoenmakers_DD}
{Schoenmakers} A.~P.,  {de Bruyn} A.~G.,  {R{\"o}ttgering} H.~J.~A.,  {van der
  Laan} H.,   {Kaiser} C.~R.,  2000b, \mn@doi [\mnras]
  {10.1046/j.1365-8711.2000.03430.x}, \href
  {http://adsabs.harvard.edu/abs/2000MNRAS.315..371S} {315, 371}

\bibitem[\protect\citeauthoryear{{Shabala} \& {Godfrey}}{{Shabala} \&
  {Godfrey}}{2013}]{shabala2013}
{Shabala} S.~S.,  {Godfrey} L.~E.~H.,  2013, \mn@doi [\apj]
  {10.1088/0004-637X/769/2/129}, \href
  {http://adsabs.harvard.edu/abs/2013ApJ...769..129S} {769, 129}

\bibitem[\protect\citeauthoryear{{Shakura} \& {Sunyaev}}{{Shakura} \&
  {Sunyaev}}{1973}]{ss1973}
{Shakura} N.~I.,  {Sunyaev} R.~A.,  1973, \aap, \href
  {http://adsabs.harvard.edu/abs/1973A%26A....24..337S} {24, 337}

\bibitem[\protect\citeauthoryear{{Sikora}}{{Sikora}}{2016}]{sikora2016}
{Sikora} M.,  2016, \mn@doi [Galaxies] {10.3390/galaxies4030012}, \href
  {http://adsabs.harvard.edu/abs/2016Galax...4...12S} {4, 12}

\bibitem[\protect\citeauthoryear{{Subrahmanyan}, {Saripalli}  \&
  {Hunstead}}{{Subrahmanyan} et~al.}{1996}]{subra1996}
{Subrahmanyan} R.,  {Saripalli} L.,   {Hunstead} R.~W.,  1996, \mn@doi [\mnras]
  {10.1093/mnras/279.1.257}, \href
  {http://adsabs.harvard.edu/abs/1996MNRAS.279..257S} {279, 257}

\bibitem[\protect\citeauthoryear{{Tadhunter}}{{Tadhunter}}{2016}]{tadhunter2016}
{Tadhunter} C.,  2016, \mn@doi [\aapr] {10.1007/s00159-016-0094-x}, \href
  {http://adsabs.harvard.edu/abs/2016A%26ARv..24...10T} {24, 10}

\bibitem[\protect\citeauthoryear{{Tazaki}, {Ueda}, {Terashima}, {Mushotzky}  \&
  {Tombesi}}{{Tazaki} et~al.}{2013}]{tazaki2013}
{Tazaki} F.,  {Ueda} Y.,  {Terashima} Y.,  {Mushotzky} R.~F.,   {Tombesi} F.,
  2013, \mn@doi [\apj] {10.1088/0004-637X/772/1/38}, \href
  {http://adsabs.harvard.edu/abs/2013ApJ...772...38T} {772, 38}

\bibitem[\protect\citeauthoryear{{Turner}, {George}, {Nandra}  \&
  {Mushotzky}}{{Turner} et~al.}{1997}]{turner1997}
{Turner} T.~J.,  {George} I.~M.,  {Nandra} K.,   {Mushotzky} R.~F.,  1997,
  \mn@doi [\apjs] {10.1086/313053}, \href
  {http://adsabs.harvard.edu/abs/1997ApJS..113...23T} {113, 23}

\bibitem[\protect\citeauthoryear{{Ubertini} et~al.,}{{Ubertini}
  et~al.}{2003}]{ibis}
{Ubertini} P.,  et~al., 2003, \mn@doi [\aap] {10.1051/0004-6361:20031224},
  \href {http://adsabs.harvard.edu/abs/2003A%26A...411L.131U} {411, L131}

\bibitem[\protect\citeauthoryear{{Ueda} et~al.,}{{Ueda}
  et~al.}{2007}]{ueda2007}
{Ueda} Y.,  et~al., 2007, \mn@doi [\apjl] {10.1086/520576}, \href
  {http://adsabs.harvard.edu/abs/2007ApJ...664L..79U} {664, L79}

\bibitem[\protect\citeauthoryear{{Urry} \& {Padovani}}{{Urry} \&
  {Padovani}}{1995}]{urry&padovani}
{Urry} C.~M.,  {Padovani} P.,  1995, \mn@doi [\pasp] {10.1086/133630}, \href
  {http://adsabs.harvard.edu/abs/1995PASP..107..803U} {107, 803}

\bibitem[\protect\citeauthoryear{{Ursini}, {Bassani}, {Panessa}, {Bazzano},
  {Bird}, {Malizia}  \& {Ubertini}}{{Ursini} et~al.}{2018a}]{ctrg}
{Ursini} F.,  {Bassani} L.,  {Panessa} F.,  {Bazzano} A.,  {Bird} A.~J.,
  {Malizia} A.,   {Ubertini} P.,  2018a, \mn@doi [\mnras]
  {10.1093/mnras/stx3159}, \href
  {http://adsabs.harvard.edu/abs/2018MNRAS.474.5684U} {474, 5684}

\bibitem[\protect\citeauthoryear{{Ursini} et~al.,}{{Ursini}
  et~al.}{2018b}]{3c382}
{Ursini} F.,  et~al., 2018b, \mn@doi [\mnras] {10.1093/mnras/sty1258}, \href
  {http://adsabs.harvard.edu/abs/2018MNRAS.478.2663U} {478, 2663}

\bibitem[\protect\citeauthoryear{{Vasudevan} \& {Fabian}}{{Vasudevan} \&
  {Fabian}}{2007}]{VF2007}
{Vasudevan} R.~V.,  {Fabian} A.~C.,  2007, \mn@doi [\mnras]
  {10.1111/j.1365-2966.2007.12328.x}, \href
  {http://adsabs.harvard.edu/abs/2007MNRAS.381.1235V} {381, 1235}

\bibitem[\protect\citeauthoryear{{Vasudevan} \& {Fabian}}{{Vasudevan} \&
  {Fabian}}{2009}]{VF2009}
{Vasudevan} R.~V.,  {Fabian} A.~C.,  2009, \mn@doi [\mnras]
  {10.1111/j.1365-2966.2008.14108.x}, \href
  {http://adsabs.harvard.edu/abs/2009MNRAS.392.1124V} {392, 1124}

\bibitem[\protect\citeauthoryear{{Verner}, {Ferland}, {Korista}  \&
  {Yakovlev}}{{Verner} et~al.}{1996}]{vern}
{Verner} D.~A.,  {Ferland} G.~J.,  {Korista} K.~T.,   {Yakovlev} D.~G.,  1996,
  \mn@doi [\apj] {10.1086/177435}, \href
  {http://adsabs.harvard.edu/abs/1996ApJ...465..487V} {465, 487}

\bibitem[\protect\citeauthoryear{{V{\'e}ron-Cetty} \&
  {V{\'e}ron}}{{V{\'e}ron-Cetty} \& {V{\'e}ron}}{2006}]{veron2006}
{V{\'e}ron-Cetty} M.-P.,  {V{\'e}ron} P.,  2006, \mn@doi [\aap]
  {10.1051/0004-6361:20065177}, \href
  {http://adsabs.harvard.edu/abs/2006A%26A...455..773V} {455, 773}

\bibitem[\protect\citeauthoryear{{Wall} \& {Peacock}}{{Wall} \&
  {Peacock}}{1985}]{2Jy_1}
{Wall} J.~V.,  {Peacock} J.~A.,  1985, \mn@doi [\mnras]
  {10.1093/mnras/216.2.173}, \href
  {http://adsabs.harvard.edu/abs/1985MNRAS.216..173W} {216, 173}

\bibitem[\protect\citeauthoryear{{Walton}, {Nardini}, {Fabian}, {Gallo}  \&
  {Reis}}{{Walton} et~al.}{2013}]{walton2013}
{Walton} D.~J.,  {Nardini} E.,  {Fabian} A.~C.,  {Gallo} L.~C.,   {Reis} R.~C.,
   2013, \mn@doi [\mnras] {10.1093/mnras/sts227}, \href
  {http://adsabs.harvard.edu/abs/2013MNRAS.428.2901W} {428, 2901}

\bibitem[\protect\citeauthoryear{{We{\.z}gowiec}, {Jamrozy}  \&
  {Mack}}{{We{\.z}gowiec} et~al.}{2016}]{wez2016}
{We{\.z}gowiec} M.,  {Jamrozy} M.,   {Mack} K.-H.,  2016, \actaa, \href
  {http://adsabs.harvard.edu/abs/2016AcA....66...85W} {66, 85}

\bibitem[\protect\citeauthoryear{{Willott}, {Rawlings}, {Blundell}  \&
  {Lacy}}{{Willott} et~al.}{1999}]{willott1999}
{Willott} C.~J.,  {Rawlings} S.,  {Blundell} K.~M.,   {Lacy} M.,  1999, \mn@doi
  [\mnras] {10.1046/j.1365-8711.1999.02907.x}, \href
  {http://adsabs.harvard.edu/abs/1999MNRAS.309.1017W} {309, 1017}

\bibitem[\protect\citeauthoryear{{Winter}, {Mushotzky}, {Tueller}  \&
  {Markwardt}}{{Winter} et~al.}{2008}]{winter2008}
{Winter} L.~M.,  {Mushotzky} R.~F.,  {Tueller} J.,   {Markwardt} C.,  2008,
  \mn@doi [\apj] {10.1086/525274}, \href
  {http://adsabs.harvard.edu/abs/2008ApJ...674..686W} {674, 686}

\bibitem[\protect\citeauthoryear{{Wozniak}, {Zdziarski}, {Smith}, {Madejski}
  \& {Johnson}}{{Wozniak} et~al.}{1998}]{wozniak1998}
{Wozniak} P.~R.,  {Zdziarski} A.~A.,  {Smith} D.,  {Madejski} G.~M.,
  {Johnson} W.~N.,  1998, \mn@doi [\mnras] {10.1046/j.1365-8711.1998.01831.x},
  \href {http://adsabs.harvard.edu/abs/1998MNRAS.299..449W} {299, 449}

\bibitem[\protect\citeauthoryear{{Zdziarski}, {Sikora}, {Pjanka}  \&
  {Tchekhovskoy}}{{Zdziarski} et~al.}{2015}]{zdziarski2015}
{Zdziarski} A.~A.,  {Sikora} M.,  {Pjanka} P.,   {Tchekhovskoy} A.,  2015,
  \mn@doi [\mnras] {10.1093/mnras/stv986}, \href
  {http://adsabs.harvard.edu/abs/2015MNRAS.451..927Z} {451, 927}

\bibitem[\protect\citeauthoryear{{van Velzen} \& {Falcke}}{{van Velzen} \&
  {Falcke}}{2013}]{VV2013}
{van Velzen} S.,  {Falcke} H.,  2013, \mn@doi [\aap]
  {10.1051/0004-6361/201322127}, \href
  {http://adsabs.harvard.edu/abs/2013A%26A...557L...7V} {557, L7}

\bibitem[\protect\citeauthoryear{{van Velzen}, {Falcke}  \& {K{\"o}rding}}{{van
  Velzen} et~al.}{2015}]{vanvelzen2015}
{van Velzen} S.,  {Falcke} H.,   {K{\"o}rding} E.,  2015, \mn@doi [\mnras]
  {10.1093/mnras/stu2213}, \href
  {http://adsabs.harvard.edu/abs/2015MNRAS.446.2985V} {446, 2985}

\makeatother
\end{thebibliography}

\end{document}